\documentclass[conference]{IEEEtran}
\IEEEoverridecommandlockouts
\usepackage{cite}
\usepackage{amsmath,amssymb,amsfonts}
\usepackage{algorithmic}
\usepackage{graphicx}
\usepackage{textcomp}
\usepackage{xcolor}
\usepackage{booktabs} 
\usepackage{bm}
\usepackage{amsmath}
\usepackage{flushend}
\usepackage{amssymb}
\usepackage{setspace}
\usepackage{cases}
\usepackage{enumitem}
\usepackage{varwidth}
\usepackage{makecell}
\usepackage{url}
\usepackage{comment}
\usepackage{array}
\usepackage[ruled]{algorithm2e} 
\usepackage{graphicx}
\usepackage[caption=false]{subfig}
\usepackage{multirow}
\usepackage{lipsum}
\usepackage{capt-of}
\usepackage{epstopdf}
\usepackage{comment}
\usepackage{authblk}
\usepackage{color, colortbl}
\usepackage[normalem]{ulem}
\usepackage{setspace}

\newcommand{\n}{GRAFICS} 
\newcommand{\linep}{E-LINE}
\newcommand{\phc}{Prox}

\definecolor{Gray}{gray}{0.9}
\allowdisplaybreaks

\begin{document}

\title{\n{}: Graph Embedding-based Floor Identification Using Crowdsourced RF Signals \vspace{-0.1in}}

\author[$\ast$]{Weipeng Zhuo}
\author[$\ast$]{Ziqi Zhao}
\author[$\ast$]{Ka Ho Chiu}
\author[$\ddagger$]{Shiju Li}
\author[$\dagger$]{Sangtae Ha}
\author[$\ddagger,\S$]{Chul-Ho Lee}
\author[$\ast,\S$]{S.-H. Gary Chan\thanks{$^\S$Corresponding authors.}}
\affil[$\ast$]{The Hong Kong University of Science and Technology}
\affil[ ]{\textsuperscript{$\dagger$}University of Colorado Boulder, \textsuperscript{$\ddagger$}Texas State University}
\affil[ ]{Email: \textsuperscript{$\ast$}\{wzhuo,zzhaoas,khchiuac,gchan\}@ust.hk, \textsuperscript{$\dagger$}sangtae.ha@colorado.edu, \textsuperscript{$\ddagger$}\{shiju.li,chulho.lee\}@txstate.edu \vspace{-0.1in}
}
\maketitle

\begin{abstract}

We study the problem of floor identification for radiofrequency (RF) signal samples obtained in a crowdsourced manner, where the signal samples are highly  heterogeneous and most samples lack their floor labels. We propose GRAFICS, a graph embedding-based floor identification system. GRAFICS first builds a highly versatile bipartite graph model, having APs on one side and signal samples on the other. GRAFICS then learns the low-dimensional embeddings of signal samples via a novel graph embedding algorithm named E-LINE. GRAFICS finally clusters the node embeddings along with the embeddings of a few labeled samples through a proximity-based hierarchical clustering, which eases the floor identification of every new sample. We validate the effectiveness of GRAFICS based on two large-scale datasets that contain RF signal records from 204 buildings in Hangzhou, China, and five buildings in Hong Kong. Our experiment results show that GRAFICS achieves highly accurate prediction performance with only a few labeled samples (96\% in both micro- and macro-$F$ scores) and significantly outperforms several state-of-the-art algorithms (by about 45\% improvement in micro-$F$ score and 53\% in macro-$F$ score).

\end{abstract}

\section{Introduction}
\label{sec:intro}
While there are plenty of smart city applications based upon radiofrequency (RF) signals, e.g., WiFi, iBeacon and UWB, their success largely depends on the availability of \emph{floor information} associated with the RF signals. For example, to facilitate pedestrian navigation in multi-floor indoor environments, RF signal-based localization systems~\cite{shokry2017tale,yang2021multi,he2016chameleon,he2015calibration} need to estimate the floor number accurately before localizing people in a two-dimensional plane. Similar situations happen with indoor unmanned aerial vehicles for scene construction or building quality monitoring~\cite{mccabe2017roles,queralta2020uwb}. In geofencing for elderly care or pandemic control~\cite{SignatureHome,monir2021iot}, RF signals are leveraged to assure that people stay on a \emph{certain} floor due to the prevalence of the RF signals. Even in robot rescue~\cite{spurny2021autonomous}, RF signals can be used in inferring the floor information when the visual information is lost.

On the other hand, crowdsourcing has emerged as a practically viable solution for the large-scale collection of data, which is crucial for the successful deployment of the above applications in practice. However, the participatory nature of crowdsourcing often makes the collected dataset incomplete. In other words, the crowdsourced RF measurement dataset contains many measurement samples lacking the exact floor information, i.e., which floor RF signals were measured. While the activities such as in-shop contactless shopping~\cite{inshop-qr} and QR code check-ins~\cite{qrcode-uk} enable the collection of `floor-labeled' RF signals (or labeled RF signals on each floor), their portion is still relatively \emph{small} in the entire dataset. To sum up, there is a need for an accurate floor identification system based on RF signals, only a few of which have their floor information.

It is, however, non-trivial to build such a floor identification system due to the following technical challenges. In the crowdsourced data collection, different people would contribute different measurements of RF signals, each of which is in the form of a vector of the pairs of detected access points (APs), more precisely, detected medium access control addresses (or MACs for short), and their corresponding received signal strength (RSS) values. The dataset is, however, \emph{heterogeneous}. The MACs observed in one sample may not appear in another. The MACs and their corresponding RSS values in the samples may also vary even when they are measured on the same floor or the same spot. This can be worse due to environmental changes (e.g., installation and removal of APs/MACs) and device heterogeneity.

\begin{figure}[t]
    \centering
    	\subfloat[]{%
        \includegraphics[width=0.45\linewidth]{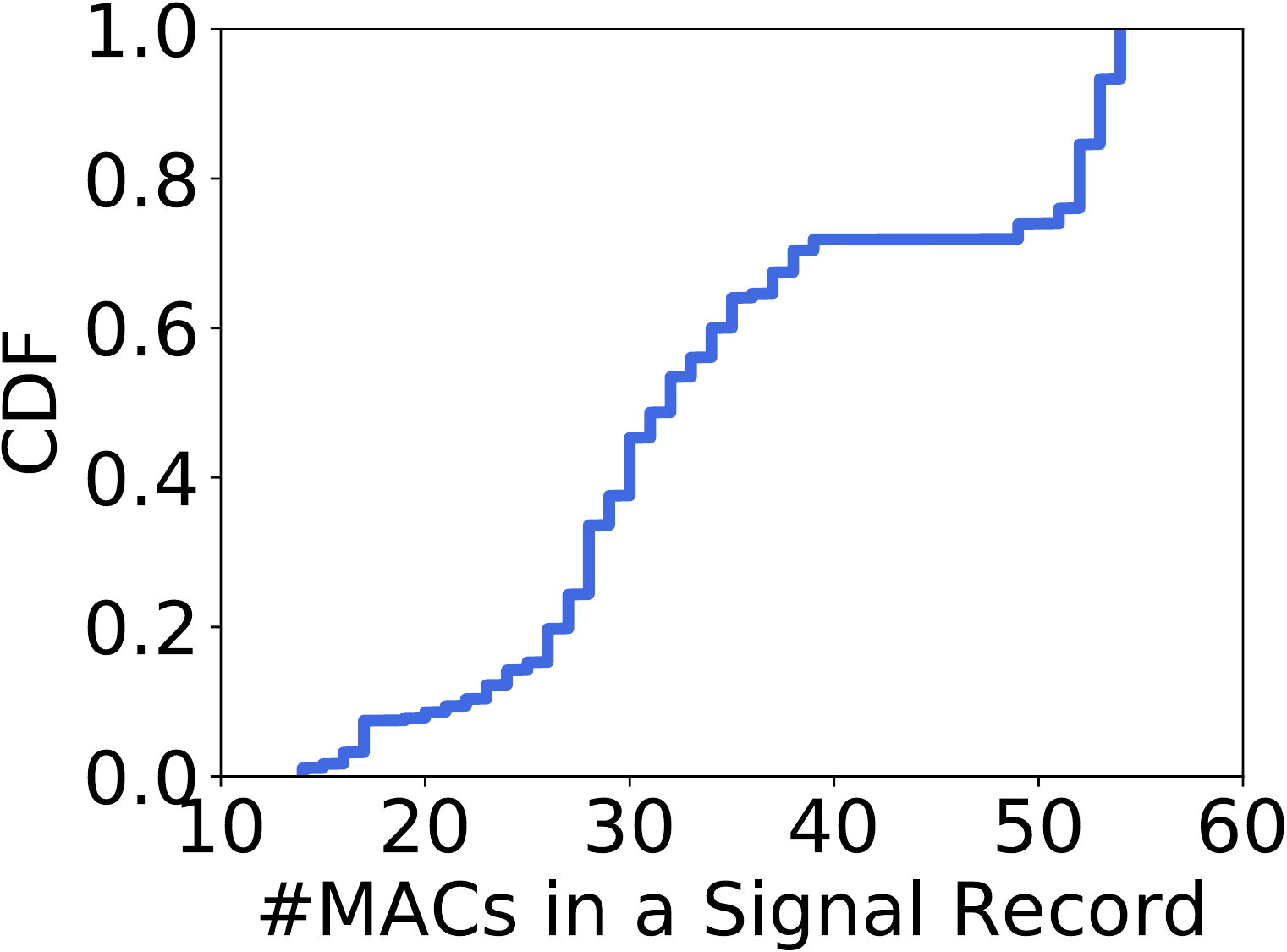}
        }
        \hspace{0.1in}
        \subfloat[]{%
        \includegraphics[width=0.45\linewidth]{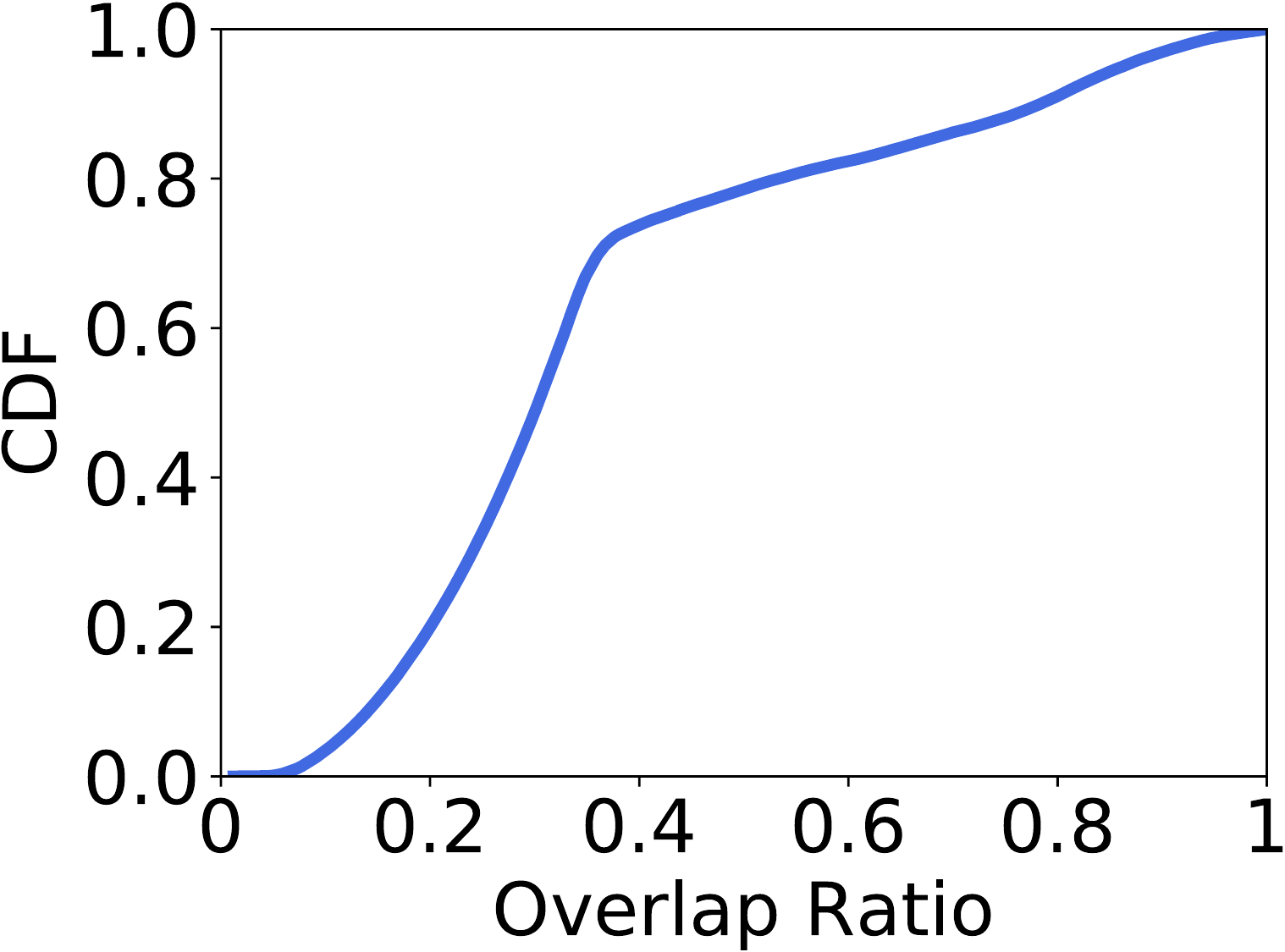}
        }
    	\caption{Statistics of RF signal records on a floor. (a) CDF of the number of MACs in a signal record; (b) CDF of the fraction of common MACs (overlap ratio) for a pair of signal records.}
    	\vspace{-0.2in}
    	\label{fig:mac_distribution}
\end{figure}

To illustrate the heterogeneity in the crowdsourced RF measurement dataset, in Figure~\ref{fig:mac_distribution}, we show the statistics of 8,274 signal records collected on a floor of a mall, where there are 805 distinct MACs. Figure~\ref{fig:mac_distribution}(a) presents the CDF of the number of MACs in each record. Most records contain less than 40 MACs. It indicates that each record only contains a small fraction of the MACs detectable on the floor. In addition, for any pair of signal records, we compute their overlap ratio, defined as the intersection over the union of MACs in the records, and plot the CDF of the overlap ratio in Figure~\ref{fig:mac_distribution}(b). Most pairs (78\%) overlap less than half. Hence, it would be \emph{inefficient} to use the dataset \emph{as is}.

In this paper, we present \n{}, {\bf gra}ph embedding-based {\bf f}loor {\bf i}dentification using {\bf c}rowdsourced RF {\bf s}ignals, which infers the floor number a user is located from an online measurement of RF signals, along with the crowdsourced RF measurement dataset. While this crowdsourced dataset can be leveraged to build a learning model for the problem, it needs to be judiciously used due to the aforementioned technical challenges. Since the floor-labeled RF measurement samples are scarce in the dataset, supervised learning models are deemed impractical, despite that the supervised learning models have been used for the floor identification~\cite{zhang2018floor,hsieh2018towards,song2019novel}. It also remains questionable how to extract useful information from the highly heterogeneous dataset. Thus, \n{} is designed to be a semi-supervised learning-based system, which is built upon graph embedding and proximity-based hierarchical clustering.

\n{} first constructs \emph{a bipartite graph} reflecting the observation in each measurement sample while being adaptive to the arrivals (or departures) of new (or old) MACs and arrivals of new measurements. In this graph, we use a node of one type to represent a MAC and a node of another type to indicate a measurement record itself. Since each measurement record has a list of observed MACs and their corresponding RSS values, we create edges between an `RF-record node' that represents a record and `MAC nodes' that correspond to the MACs observed in the record. We then assign weight values to the edges based on the RSS values. The connectivity structure in the graph captures the relevance between measurement records while reflecting the information in each record.

To cope with the heterogeneity in the RF samples, \n{} next learns \emph{a low-dimensional vector representation} (or embedding) of each node in the graph. As a byproduct, \n{} prevents the so-called missing value problem in the first place, which arises when the measurement samples are directly used in a matrix form~\cite{nowicki2017low, alitaleshi2020wifi} (see Section~\ref{sec:related} for more details), since the node embeddings of equal length are used on behalf of the measurement samples whose dimensions can be different. To obtain the node embeddings, we adopt LINE~\cite{LINE}, yet with our own refinements.

While the original LINE algorithm learns node embeddings based on the relationship between nodes via their common neighbors, we extend the algorithm to learn the embeddings to capture the relationship between nodes based on not only their common `direct' neighbors but also other common `multi-hop' neighbors. Our extension of LINE, named \linep{}, is better suited for learning the node embeddings of our bipartite graph for floor classification. For example, two RF measurement records, even though they are taken on the same floor, may not share many common MACs but contain the MACs that rather overlap with the ones in the other records collected on the same floor. We empirically demonstrate that \n{} (with \linep{}) indeed outperforms its version with the LINE algorithm for the floor classification problem (see Section~\ref{sec:exp}).

Once the node embeddings are obtained, \n{} uses a proximity-based hierarchical clustering to cluster measurement samples in the embedding space so that each cluster forms around each floor-labeled sample, i.e., an RF measurement sample that comes with its floor information. Thanks to graph embedding and hierarchical clustering, \n{} is able to work with \emph{much fewer} floor-labeled samples, when compared to the prior supervised learning-based solutions~\cite{zhang2018floor,hsieh2018towards,song2019novel} (see Section~\ref{sec:exp}). Finally, whenever there is an \emph{online} measurement of RF signals for \emph{inference} (i.e., identifying which floor the measurement is taken), its embedding is first obtained through the graph embedding, and its nearest cluster is then identified based on its distance to the centroid of each cluster in the embedding space. The floor associated with the nearest cluster eventually becomes the inference result.

Our contributions can be summarized as follows.

\begin{itemize}[itemsep=3pt,leftmargin=1.5em]

\item \emph{\n{} is capable of handling the heterogeneity in the RF samples}. The construction of a bipartite graph incorporates both the information in each RF measurement record and the relationship between the records. In addition, the graph is easily extendable for new RF records. It can also be adjusted to reflect installation and removal of APs.

\item \emph{\n{} learns the representations of RF samples}.
    Our novel graph embedding algorithm \linep{} maps each node in the bipartite graph onto the same low-dimensional vector space, which enables a better identification of similarities and differences between the heterogeneous RF samples. \linep{} also produces \emph{better} representations or node embeddings in improving the performance of floor identification, as compared to LINE.

\item \emph{\n{} requires a small amount of labeled data.} The node embeddings are used to cluster RF samples (including few floor-labeled samples) in the embedding space via the hierarchical clustering. This semi-supervised learning makes \n{} require \emph{much fewer} floor-labeled samples than the supervised ones. This efficiency stems from the \emph{enhanced} vector representations of the RF samples.

\item \emph{\n{} significantly outperforms state-of-the-art methods.} We implement \n{} and evaluate its performance based on Microsoft's Kaggle open dataset, which includes crowdsourced RF samples obtained in 204 multi-floor buildings in Hangzhou, China, and a newly collected dataset, which has RF samples obtained in 5 buildings in Hong Kong. Experimental results show that \n{} achieves higher than 96\% in micro-$F$ and macro-$F$ scores across all buildings with only few labels per floor (around 4 out of 1000 on average). Compared with state-of-the-art methods, \n{} also outperforms around 45\% in micro-$F$ score and 53\% in macro-$F$ score.

\end{itemize}

The rest of this paper is organized as follows. We review related work in Section~\ref{sec:related} and provide an overview of \n{} in Section~\ref{sec:system}. In Section~\ref{sec:offline}, we explain the details of bipartite graph modeling, graph embedding, and hierarchical clustering in \n{}. We then elaborate on how the online inference is done in \n{} in Section \ref{sec:online}. Extensive experimental results are presented in Section~\ref{sec:exp}. We conclude in Section~\ref{sec:conclude}.

\section{Related Work}
\label{sec:related}
\noindent \textbf{Floor classification with different sensors:} Different sensors from smartphones have been leveraged for the floor classification problem~\cite{ascher2012multi,yeFLocFloorLocalization2014,shen2015barfi}. A heuristic thresholding method is used with barometer readings~\cite{zhao2015hyfi} to detect floor transitions. However, such a threshold is difficult to set in practice since different barometer models may have different sensitivity levels. Thus, in~\cite{li2018multi}, the barometer readings are fused with WiFi and inertial measurement unit (IMU) sensors using Kalman filter to infer floor information. Magnetometers from smartphones are also explored in~\cite{ashraf2019floor} with IMU readings for floor classification. Despite their satisfactory performance, the collection of such sensor readings introduces high overhead for data storage, data transmission, and battery consumption. Furthermore, users are supposed to follow carefully designed trajectories as ground truth~\cite{chen2021joint} to collect sensor data such that they can be well `labeled'. In contrast, \n{} is \emph{lightweight} and based only on \emph{crowdsourced RF measurements} to achieve accurate and efficient detection of floor numbers.

\vspace{1mm}
\noindent \textbf{The need for massive labeled data:} Several machine learning algorithms have been utilized for floor classification~\cite{zhang2018floor,hsieh2018towards,elbakly2018truestory,dou2020bisection,elbakly2020storyteller,wang2021secure}. For example, a support vector machine-based method is proposed in~\cite{zhang2018floor}, but it needs to train support vectors for the classification of every pair of floors, which is inconvenient in practice. To address this problem, a recurrent neural network approach is introduced in~\cite{hsieh2018towards}. It works well when data has a temporal relationship, which implies that one would need to obtain a group of RF samples collected from the same trajectory. However, crowdsourced RF samples are usually sporadic and contributed by many different users. Furthermore, all these machine learning-based solutions require a large amount of labeled data in training their classification model. In contrast, \n{} works with only \emph{few} labeled samples.

\vspace{1mm}
\noindent \textbf{Leveraging the locations of APs:} There are also a few prior studies~\cite{elbakly2018truestory,caso2019vifi,elbakly2020storyteller} that leverage the knowledge of APs' locations for floor classification. For instance, ViFi~\cite{caso2019vifi} learns the parameters of a signal propagation model from the RSS measurements to generate virtual reference points and then predicts floor labels for new signals with a weighted $k$-nearest neighbors algorithm. StoryTeller~\cite{elbakly2020storyteller} converts RF signals to images based on APs with strong signal strengths and then trains a convolutional neural network (CNN) model for floor classification. They both require the APs' locations, which are, however, usually unavailable when the RF measurement data is collected in a crowdsourced manner. Thus, \n{} is designed to be \emph{independent} of the AP locations and only uses the signal readings collected from different APs to infer floor information accurately.

\vspace{1mm}
\noindent \textbf{Assuming fixed-length RF signal vectors:} When using RF signals for floor classification, the state-of-the-art systems~\cite{nowicki2017low,kim2018scalable,song2019novel,alitaleshi2020wifi} first group all the RF signal data together in a matrix form. Each RF signal measurement record contains a set of RSS values from surrounding APs that are represented by their MAC addresses, so the RSS-value set forms a row of the matrix. In other words, each sensed MAC address forms a column where the RSS value associated with the MAC address in each record is the value of an entry. Once the matrix is constructed, they transform each row in the matrix to a low-dimensional vector space via a CNN or an autoencoder, and build a classification model for floor identification.

\begin{figure}[t]
	\centering
	\includegraphics[width=0.4\textwidth]{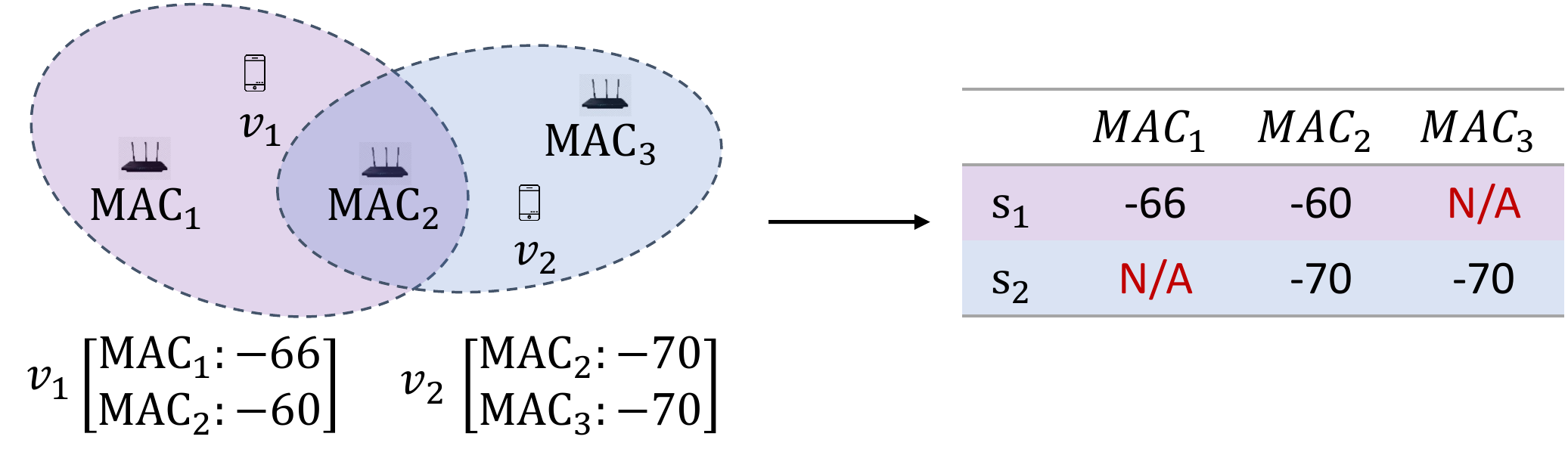}
	\vspace{-0.1in}
	\caption{Missing value problem when representing RF signal data in a matrix.}
	\label{fig:matrix_form}
	\vspace{-0.2in}
\end{figure}

In the matrix representation, however, there is a so-called missing value problem in that some entries in the matrix lack RSS values since the RF samples are of variable length (i.e., the number of detected MAC addresses can be different per sample), as shown in Figure~\ref{fig:matrix_form}. The missing entries are generally filled with extremely small values to indicate nonavailability. Nonetheless, some missing values may be due to the limited scanning capability of low-end devices. The entries with the extremely small values could also be misinterpreted as the presence of their corresponding APs, albeit with weak signals, when used for building a classifier. Thus, such ad-hoc data imputation could potentially lead to erroneous feature extraction, resulting in unsatisfactory floor classification and learning inefficiency. In contrast, \n{} models RF signal records using a bipartite graph, which naturally \emph{avoids} the missing value problem.

\section{\n{}: Design}
\label{sec:system}
\begin{figure*}[t]
    \centering
    \subfloat[Offline Training]{%
        \includegraphics[width=0.58\linewidth]{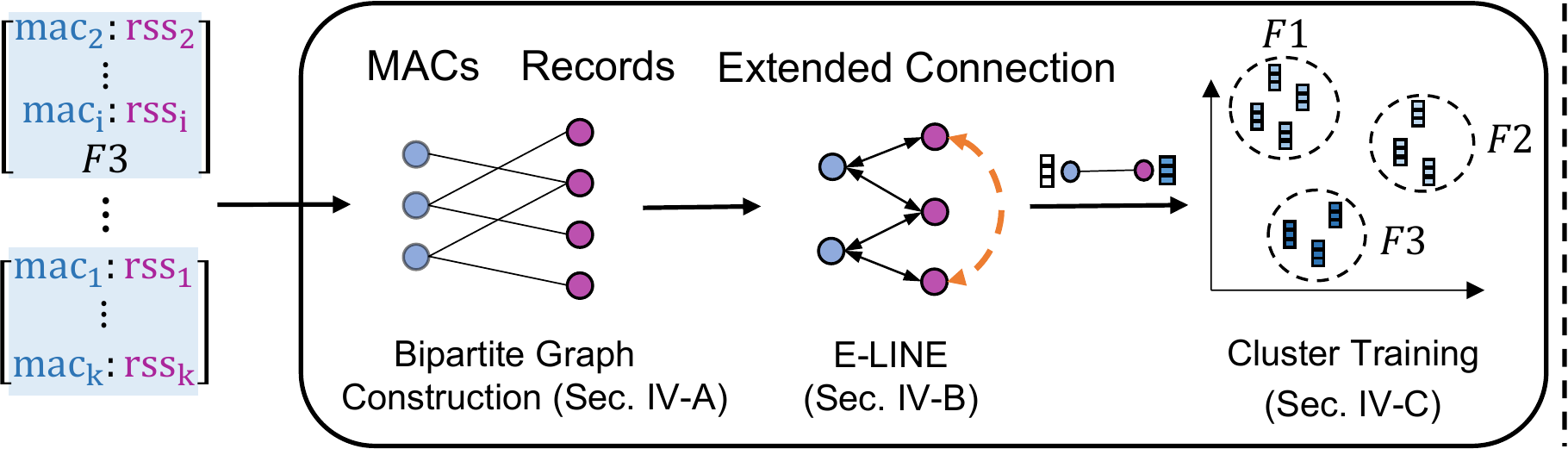}
        }
    \subfloat[Online Inference]{%
        \includegraphics[width=0.39\linewidth]{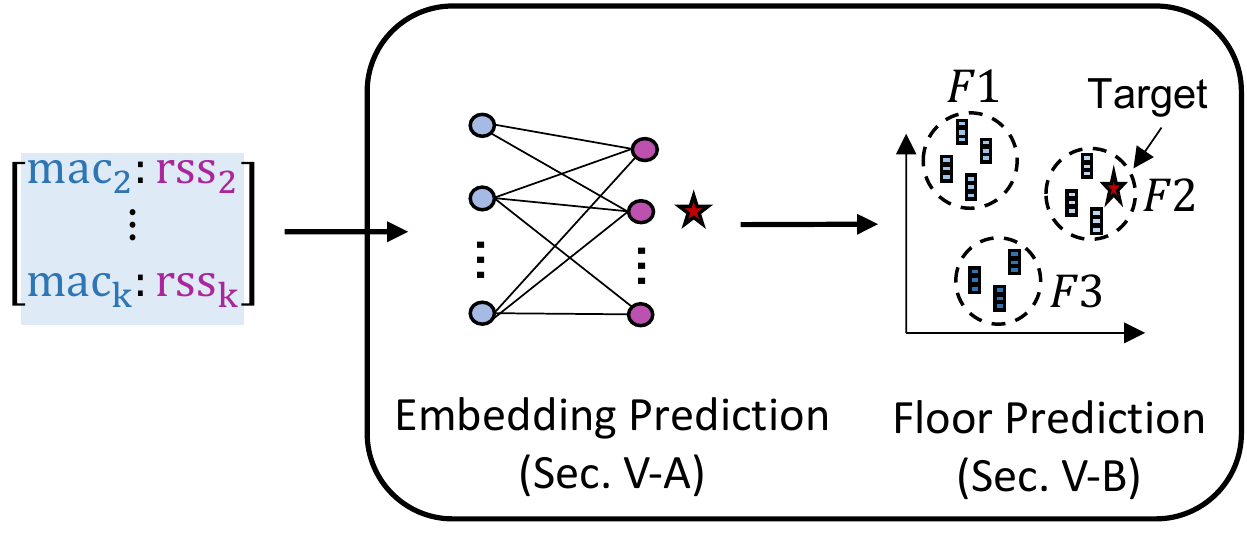}
        }
    \caption{A system overview of \n{}.}
    	\label{fig:sys_overview}
    	\vspace{-0.2in}
\end{figure*}

We first discuss the technical challenges that are considered and addressed in designing our floor identification system \n{}. We then provide a system overview before we go into the details in the subsequent sections.

\subsection{Technical Challenges and Considerations}
\label{subsec:design_challenge}

\noindent \textbf{Dynamic RF environments:} Crowdsourced RF signal samples consist of scattered, noisy RSS values and are of variable sizes. For example, the signal samples collected at two different places yet on the same floor may have no overlap with each other because of the limited coverage of APs. Even on the same spot, the recorded RSS values may differ and detected APs may vary due to environmental changes and device heterogeneity. The RSS records could also vary over time, and APs could be added and removed over time. To address these issues, we propose a novel representation of the scattered, noisy RF signal samples based on a bipartite graph, which allows us to uncover the hidden relationship among the RF samples. We also propose \linep{} that improves on LINE in a way that is better suitable for our bipartite graph to map each node (an RF sample) in the graph into a low-dimensional vector space. Since the RF samples (of variable length) are now represented in the same embedding space, their similarities and differences can be easily measured via the distances among the samples in the embedding space.

\vspace{1mm}
\noindent \textbf{Scarce floor-labeled data:} Crowdsourcing enables a large-scale collection of RF signal data but does not ensure the collection of their contextual information such as which floor they were obtained. This floor information is the \emph{label} information in the floor identification problem. While we can still collect the floor-labeled RF samples from the events such as in-shop contactless shopping~\cite{inshop-qr} and QR code check-ins~\cite{qrcode-uk}, their portion in the dataset is largely limited. In other words, not every RF sample is associated with its floor label, and only a very small portion of the RF data has the floor labels. Thus, after obtaining the embeddings of the RF samples, we do not build any supervised learning model for floor classification, which often requires a considerable amount of labeled data. Instead, thanks to the ease of computing the distances among the samples via their embeddings, we are able to build an effective semi-supervised learning model with only a tiny fraction of labeled data, which is done via our proposed proximity-based hierarchical clustering.

\subsection{System Overview}
\label{subsec:sys_overview}

Figure~\ref{fig:sys_overview} illustrates an end-to-end workflow of \n{}, which ranges from offline training to online inference. In the offline training phase, \n{} first models RF signal measurement samples using a bipartite graph, inspired by their structure where each RF signal sample consists of sensed MACs with corresponding RSS values (Section~\ref{subsec:bipartite}). In this bipartite graph, the RF signal samples and sensed MACs are represented by two different types of nodes, respectively, and they are connected by edges weighted by the RSS values from the sensed MACs (or APs). As a result, changes in the MACs or RF signal samples can be easily captured by adding or removing the nodes or edges in the graph. Then, the bipartite graph is processed with our proposed embedding algorithm \linep{} (Section~\ref{subsec:dline}) to learn low dimensional coordinates, i.e., a latent representation or embedding for each node in an unsupervised manner. Finally, the learned representations, together with the small set of labels on each floor, are used to build a simple yet effective floor classifier via our proximity-based hierarchical clustering (Section~\ref{subsec:cluster_train}). During this process, all unlabeled crowdsourced signal samples form clusters, each of which also contains a floor-labeled sample, so their labels are virtually predicted as the label of the floor-labeled sample in the cluster that they belong to.

On the other hand, in the online \emph{inference} phase, for each newly collected RF sample, \n{} first generates its embedding through the graph embedding (Section~\ref{subsec:network_predict}) and then predicts its label (or infer which floor the new sample is taken) as the label (or floor) of the cluster that is closest in the embedding space (Section~\ref{subsec:floor_predict}).

\section{\n{}: Offline Training}
\label{sec:offline}
The offline training of \n{} is done via the following three steps: (i) constructing a bipartite graph based on the RF signal samples, (ii) learning node embeddings from the graph by our \linep{} algorithm, and (iii) building a simple yet effective classification model via our proximity-based hierarchical clustering method.

\subsection{Bipartite Graph Construction}
\label{subsec:bipartite}

Conventionally, RF signal data are represented in a matrix form (or vectors of equal length) for indoor localization and floor identification applications~\cite{nowicki2017low,kim2018large,kim2018scalable,song2019novel,alitaleshi2020wifi,abbas2019wideep}. Such a matrix representation, however, often suffers from the missing value problem, as explained in Section~\ref{sec:related}. Thus, we propose to model the variable-length RF signal records as \emph{a weighted bipartite graph}, where the measured signal information is preserved without having the missing value problem. In the graph, each RF signal record is a node of one type and the sensed MAC addresses in each record are nodes of the other type. Edges indicate the presence of sensed MAC addresses in each RF record, and their edge weights are determined based on the RSS values from the corresponding APs.

\begin{figure}[t]
	\centering
	\includegraphics[width=0.4\textwidth]{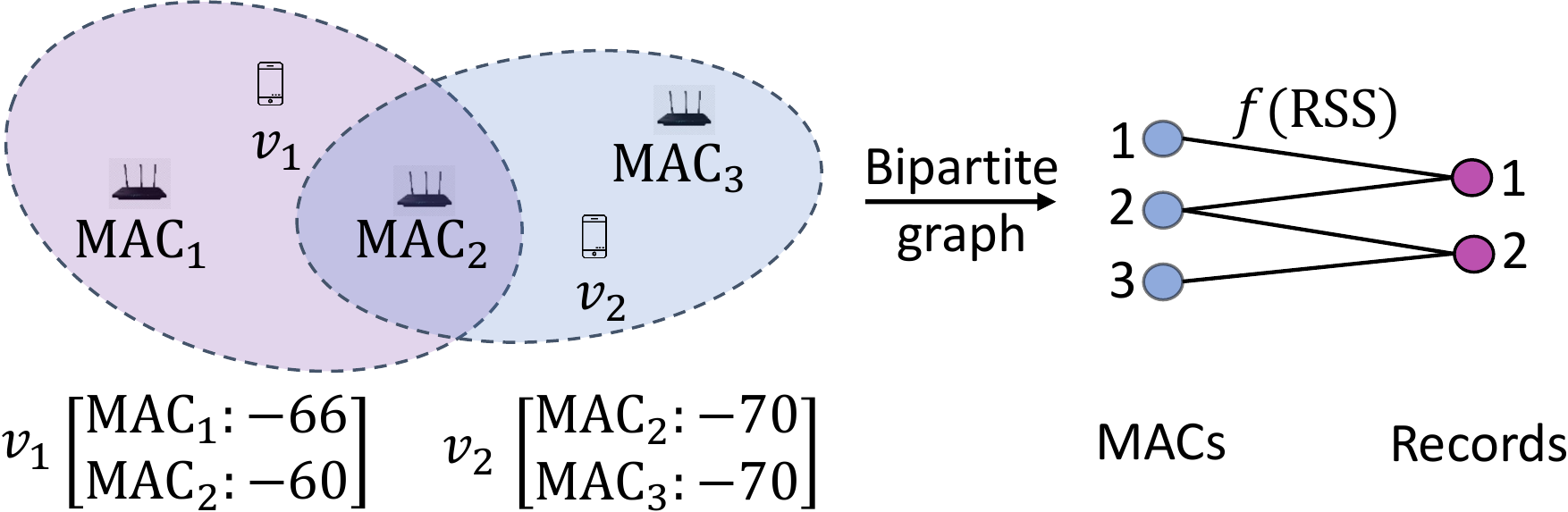}
	\caption{Example of two signal records with three sensed MACs.}
	\label{fig:bipartite}
	\vspace{-0.2in}
\end{figure}

Let $\mathcal{G} \!=\! (\mathcal{M}, \mathcal{V}, \mathcal{E})$ be a weighted bipartite graph. $\mathcal{V}$ is a set of nodes to represent the RF signal records, and $\mathcal{M}$ is a set of nodes for the sensed MAC addresses. Also, $\mathcal{E}$ is a set of edges, where $e_{mv} \in \mathcal{E}$ denotes the edge between $m \in \mathcal{M}$ and $v \in \mathcal{V}$. Note that the edge $e_{mv}$ indicates the presence of MAC address $m$ in record $v$. For each record $v$, let $\text{RSS}_{mv}$ be the RSS value from an AP whose MAC address is $m$, which appears in the record. We then associate each edge $e_{mv}$ with weight $c_{mv}$, which is defined as
\begin{equation}
c_{mv} := f(\text{RSS}_{mv}),
\label{eqn:edge_weight}
\end{equation}
where $f$ is a function of $\text{RSS}_{mv}$ with $f(\text{RSS}_{mv}) \!>\! 0$ for all $\text{RSS}_{mv}$. Note that two nodes $m$ and $v$ are connected as long as there exists a measured value of $\text{RSS}_{mv}$ from MAC $m$ in record $v$, with the edge weight being $c_{mv}$. Figure~\ref{fig:bipartite} illustrates an example where signal record $v_1$ is connected to MACs 1--2, and $v_2$ is connected to MACs 2--3.

We use the following weight function for graph $\mathcal{G}$:
\begin{equation}
f(\text{RSS}_{mv}) := \text{RSS}_{mv} + \alpha,
\label{eqn:offset}
\end{equation}
with some constant $\alpha$ that satisfies $\alpha \!>\! \max\{|\text{RSS}_{mv}|, \forall m, v\}$. In Section~\ref{subsec:exp_micro}, we empirically show that it achieves the best performance, implying that a constant offset preserves the relationships among nodes better than other schemes.

\subsection{\linep{}}
\label{subsec:dline}

From the bipartite graph $\mathcal{G}$, we next obtain vector representations or embeddings of nodes to be used to build a classifier for floor identification. To this end, we propose a novel graph embedding algorithm \linep{}, which improves on LINE~\cite{LINE} in learning better vector representations of equal length from the bipartite graph $\mathcal{G}$. Among others, LINE is a widely adopted graph embedding algorithm in the literature~\cite{ma2021deep}. Note that graph embedding here is representation learning when each node in a graph is not associated with a set of features (or attributes), and the embedding vectors are learned mainly based on the underlying connectivity structure of a given graph, unlike graph neural networks.

Before going into the details of \linep{}, we below briefly explain LINE since \linep{} is an extension of LINE. LINE considers the following two factors to learn node embeddings from a given graph: (i) whether two nodes are connected, and (ii) whether two nodes share common neighbors. These two factors lead to two different notions of `proximity', which are first-order proximity and second-order proximity, respectively. We observe that the first-order proximity is no longer useful when LINE is used for a bipartite graph. The edges only exist between the nodes of different types, but we are more interested in the relationships among the nodes of the same type. Our experiment also confirms that LINE performs better with the second-order proximity only than the one using both proximities in learning node embeddings from the bipartite graph for the floor identification problem. Thus, we below focus on the second-order proximity of LINE and explain how it is extended in \linep{}.

The second-order proximity in LINE is defined for a \emph{directed} graph. Note that any undirected graph can be interpreted as a directed graph by replacing each (undirected) edge with two directed edges having the opposite directions. The weight for each undirected edge is also used as the weights for both directed edges, i.e., $c_{ij} = c_{ji}$. The rationale behind the second-order proximity is to make nodes $i$ and $j$ close to each other in the embedding space, if they share many common neighbors, i.e., $|N(i) \cap N(j)|$ is large, where $N(i)$ and $N(j)$ denote the sets of the neighbors of $i$ and $j$, respectively.

To be precise, there are two different embeddings for each node $i$. One is `ego' embedding, defined as $\bm{u}_i$, to represent the node itself, and it is the representation of each node to use in building a floor classifier. The other is `context' embedding, defined as $\bm{u}'_i$, to represent the relationship between node $i$ and other nodes, or encode its (one-hop) neighborhood information. The second-order proximity between $i$ and $j$ is determined by the similarity between their \emph{context} embeddings $\bm{u}'_i$ and $\bm{u}'_j$. If two nodes share many common neighbors, their \emph{context} embeddings are similar to each other, so are their \emph{ego} embeddings. Here the embedding dimensions are the same and determined as a hyperparameter.

The second-order proximity is defined based on the conditional probability of the \emph{context} of $j$ determined given node $i$, which is given by
\begin{equation}
	\Pr(\bm{u}'_j|\bm{u}_i) := \frac{\exp(\bm{u}'_j \cdot \bm{u}_i)}{\sum_{l \in \mathcal{M}\cup \mathcal{V}}\exp(\bm{u}'_l \cdot \bm{u}_i)}~\label{eqn:p2},
\end{equation}
where $\bm{a} \cdot \bm{b}$ is the inner product between $\bm{a}$ and $\bm{b}$. Its empirical probability is also defined as
\begin{equation}
	\hat{\Pr}(\bm{u}'_j|\bm{u}_i) := \frac{c_{ij}}{\sum_{l \in N(i)}c_{il}}~\label{eqn:p2_hat},
\end{equation}
which characterizes the `influence' from node $i$ to the \emph{context} of its neighbor $j$. Then, the ego and context embeddings are determined so that the following objective function is minimized:
\begin{equation*}
	\mathcal{O}_1 = \sum_{i \in \mathcal{M} \cup \mathcal{V}} \lambda_i \sum_{j\in N(i)} \hat{\Pr}(\bm{u}'_j|\bm{u}_i)\log \frac{\hat{\Pr}(\bm{u}'_j|\bm{u}_i)}{\Pr(\bm{u}'_j|\bm{u}_i)},
\end{equation*}
which is defined based on the Kullback–Leibler divergence of $\Pr(\bm{u}'_j|\bm{u}_i)$ from $\hat{\Pr}(\bm{u}'_j|\bm{u}_i)$ with the convention $0 \log 0 \!=\! 0$. After omitting some constant terms, the objective function can be written as
\begin{equation*}
	\mathcal{O}_1 = - \sum_{i \in \mathcal{M} \cup \mathcal{V}} \sum_{j \in N(i)}\lambda_i \frac{c_{ij}}{\sum_{l\in N(i)}c_{il}}\log \Pr(\bm{u}'_j|\bm{u}_i).
\end{equation*}
By setting $\lambda_i := \sum_{l\in N(i)}c_{il}$, we have
\begin{equation}
	\mathcal{O}_1 = - \sum_{i \in \mathcal{M} \cup \mathcal{V}} \sum_{j \in N(i)} c_{ij}\log \Pr(\bm{u}'_j|\bm{u}_i).
	\label{eqn:objective2}
\end{equation}

We see that the second-order proximity mainly captures the closeness between nodes in the (ego) \emph{embedding} space via the similarity between their context embeddings. Since each context embedding encodes the one-hop neighborhood information of each node, their similarity indicates how many one-hop neighbors they commonly share. However, it is not able to capture the relevance between nodes when they do not share many neighbors in their one-hop neighborhood but in their local yet few-hop neighborhood. In other words, the context embeddings $\bm{u}'_i$ and $\bm{u}'_j$ of nodes $i$ and $j$ should also be similar to each other if they are reachable from each other in a few hops via multiple different local paths. This is particularly important in our floor identification problem. For example, two RF measurement samples taken on the same floor may not share many common MACs but contain the MACs that rather overlap with the ones in the other samples collected on the same floor. While the second-order proximity cannot capture such a case, their context embeddings should be similar to each other so that they can be mapped closely on the (ego) embedding space.

\begin{figure}[t]
    \centering
    \includegraphics[width=0.3\textwidth]{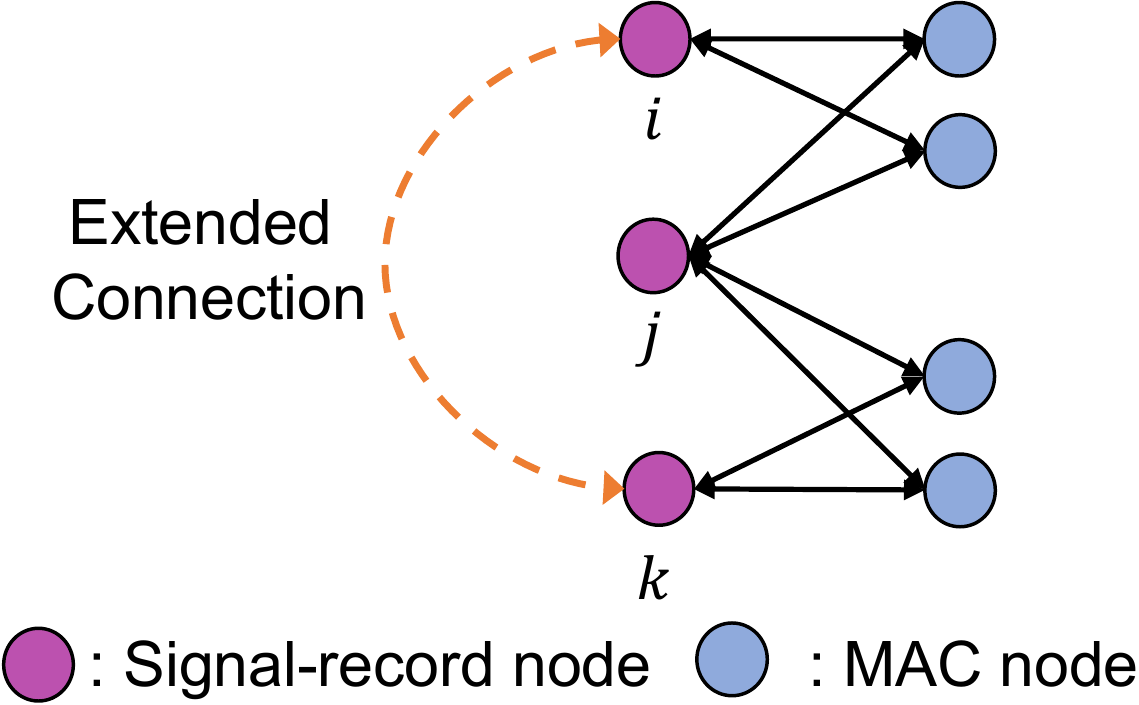}
    \caption{\linep{} incorporates indirect relationships between nodes into graph embedding.}
    \label{fig:line_plus}
    \vspace{-0.2in}
\end{figure}

To achieve better node embeddings, we below introduce a novel extension of LINE, or \linep{}. Similar to~(\ref{eqn:p2}), we define the conditional probability of the ego (node) $j$ influenced given the context of $i$, which is given by
\begin{equation}
    \Pr(\bm{u}_j|\bm{u}'_i) = \frac{\exp(\bm{u}_j \cdot \bm{u}'_i)}{\sum_{l \in \mathcal{M}\cup \mathcal{V}}\exp(\bm{u}_l \cdot \bm{u}'_i)}~\label{eqn:p3}.
\end{equation}
Its empirical distribution is also defined as
\begin{equation}
	\hat{\Pr}(\bm{u}_j|\bm{u}'_i) = \frac{c_{ij}}{\sum_{l \in N(i)}c_{il}},~\label{eqn:p3_hat}
\end{equation}
which corresponds to~(\ref{eqn:p2_hat}). Following the same line of argument as above, we can construct the following objective function:
\begin{align}
    \mathcal{O}_2 = -\sum_{i \in \mathcal{M} \cup \mathcal{V}} \sum_{j \in N(i)} c_{ij}\log \Pr(\bm{u}_j|\bm{u}'_i).
    \label{eqn:objective3}
\end{align}
Finally, we learn the ego and context embeddings by minimizing the following objective function, which is a combination of the ones in~(\ref{eqn:objective2}) and (\ref{eqn:objective3}):
\begin{align}
    &\mathcal{O}_3 = \mathcal{O}_1 + \mathcal{O}_2 \nonumber\\
    &~ = -\!\!\sum_{i \in \mathcal{M} \cup \mathcal{V}} \sum_{j \in N(i)} \! c_{ij}\left(\log \Pr(\bm{u}'_j|\bm{u}_i) \!+\!  \log \Pr(\bm{u}_j|\bm{u}_i')\right).\! \label{eqn:objective}
\end{align}
To summarize, \linep{} enhances LINE in such a way that nodes are mapped into the embedding space based on the similarity between their context embeddings that encode their \emph{local} neighborhood information, which is now more than just their one-hop neighborhood information. See Figure~\ref{fig:line_plus} for illustration, where nodes $i$ and $k$ would also be close to each other in the embedding space.

On the other hand, minimizing the objective function in (\ref{eqn:objective}) is computationally expensive in practice since the normalizing constants of $\Pr(\bm{u}'_j|\bm{u}_i)$ and $\Pr(\bm{u}_j|\bm{u}_i')$ involve the summations over the entire node set $\mathcal{M}\cup \mathcal{V}$. Thus, instead of the objective function in (\ref{eqn:objective}), we consider its relevant objective function, which is based on the widely used `negative sampling' method (see~\cite{Mikolov2013,LINE,ma2021deep} and references therein) and is computationally much easier to evaluate. Specifically, we use the following objective function to minimize in learning the embeddings of each node:
\begin{align}
    \mathcal{L}_{\mathcal{G}} := &-\sum_{i \in \mathcal{M} \cup \mathcal{V}} \sum_{j \in N(i)}c_{ij} \Big(\log \left[ \sigma (\bm{u}'_{j} \cdot \bm{u}_{i}) \sigma(\bm{u}_{j} \cdot \bm{u}'_{i}) \right]  \nonumber\\
    & + K \mathbb{E}_{z \sim \Pr(z)} \left[\log \left( \sigma(- \bm{u}'_{z} \cdot \bm{u}_{i}) \sigma (- \bm{u}_{z} \cdot \bm{u}'_{i}) \right) \right]\Big), \label{eqn:loss}
\end{align}
where $\sigma(x) \!:=\! 1/(1+\exp(-x))$. The second term in each summand is based on $K$ `negative' samples. The expectation $\mathbb{E}$ is with respect to node $z$ that is randomly drawn according to a probability distribution $\Pr(z)$, $z \!\in\! \mathcal{M} \cup \mathcal{V}$, and is computed via the Monte Carlo approximation with $K$ negative samples. As widely used in the literature~\cite{Mikolov2013,LINE,ma2021deep}, we choose $\Pr(z) \propto d_z^{3/4}$, where $d_z$ is the degree of node $z$.

Recall that the goal here is to make nodes close to each other in the (ego) embedding space if they share many common local neighbors. The rationale behind minimizing the objective function in (\ref{eqn:loss}) is as follows. On one hand, the first term in each summand makes the ego (or context) embedding of $i$ and the context (or ego) embedding of $j$ similar to each other for every pair of neighboring nodes $i$ and $j$. We can also see that due to the presence of the context embedding of each node, the ego embeddings of nodes that are in their local neighborhood would eventually be similar to each other. On the other hand, the second term in each summand, which is based on $K$ negative samples, makes the ego (or context) embedding of $i$ and the context (or ego) embedding of $z$ \emph{dissimilar} to each other. Here node $z$ is most likely to be a node \emph{outside} the local neighborhood of $i$. Therefore, the ego embeddings of locally closed nodes would be similar to each other, while the ones of distant nodes would be dissimilar to each other.

\begin{figure*}[t]
    \centering
    \subfloat[\linep{} (\n{})]{%
        \includegraphics[width=0.2\linewidth]{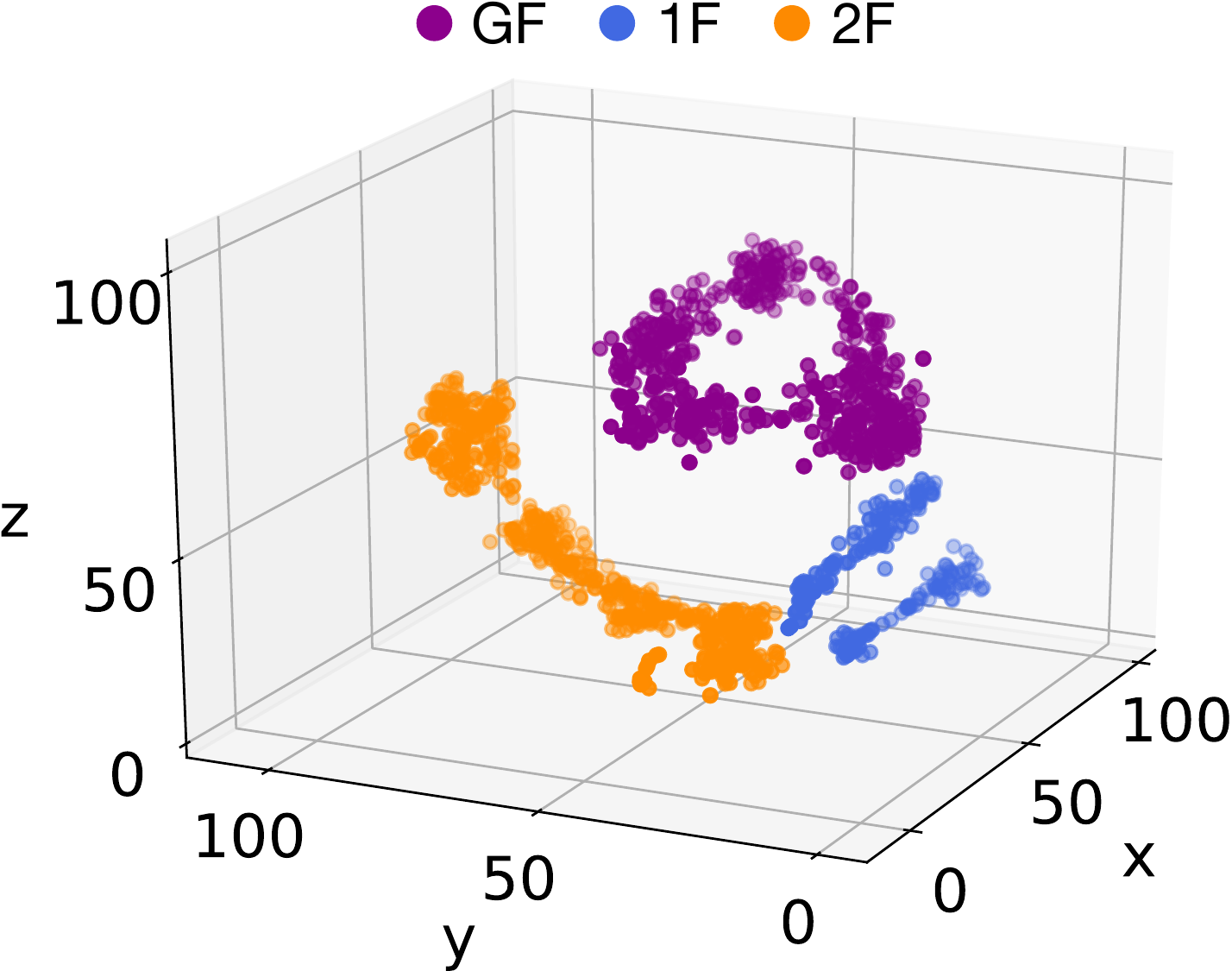}
        }
        \hspace{0.3in}
    \subfloat[MDS]{%
        \includegraphics[width=0.2\linewidth]{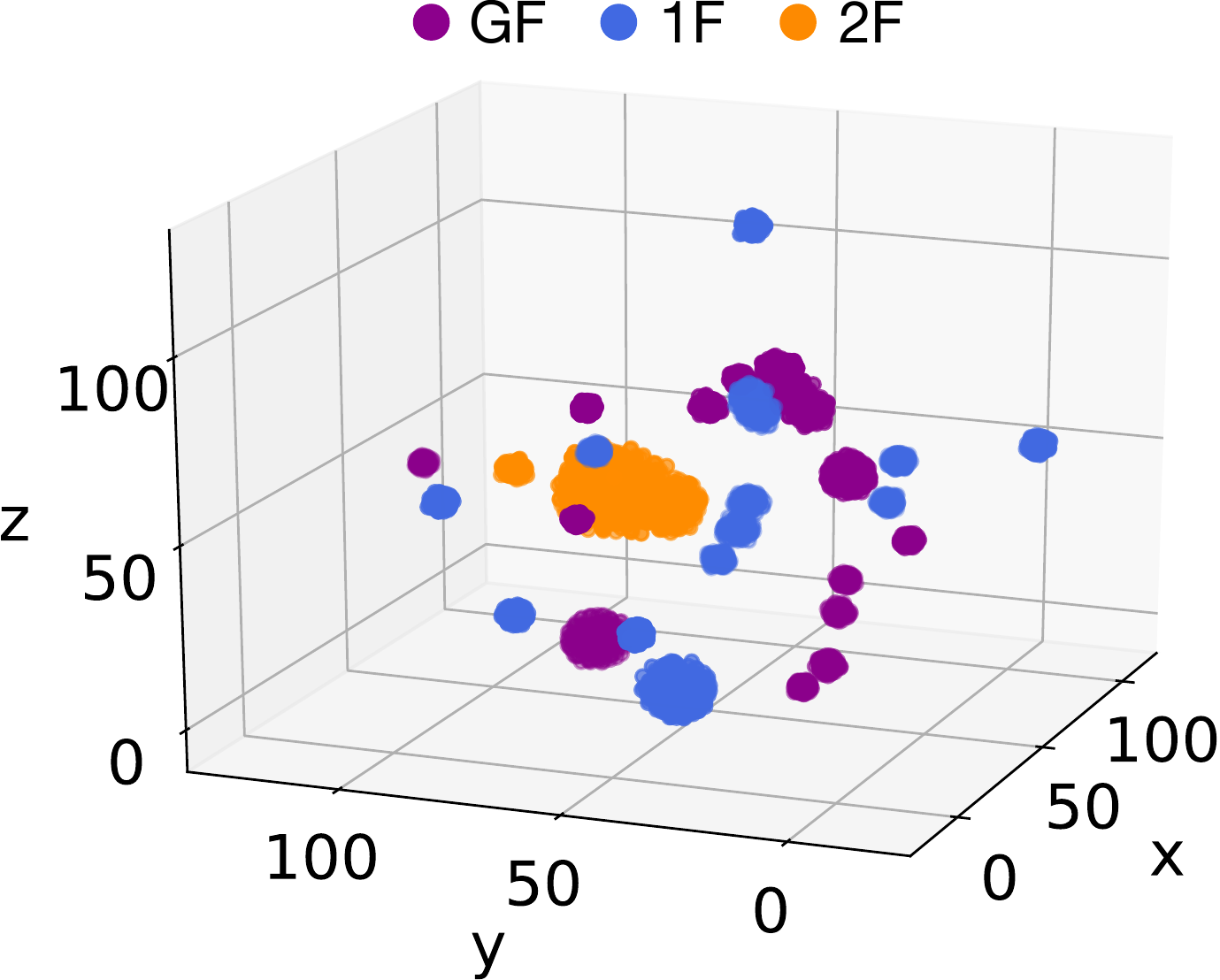}
        }
        \hspace{0.3in}
    \subfloat[Autoencoder]{%
        \includegraphics[width=0.2\linewidth]{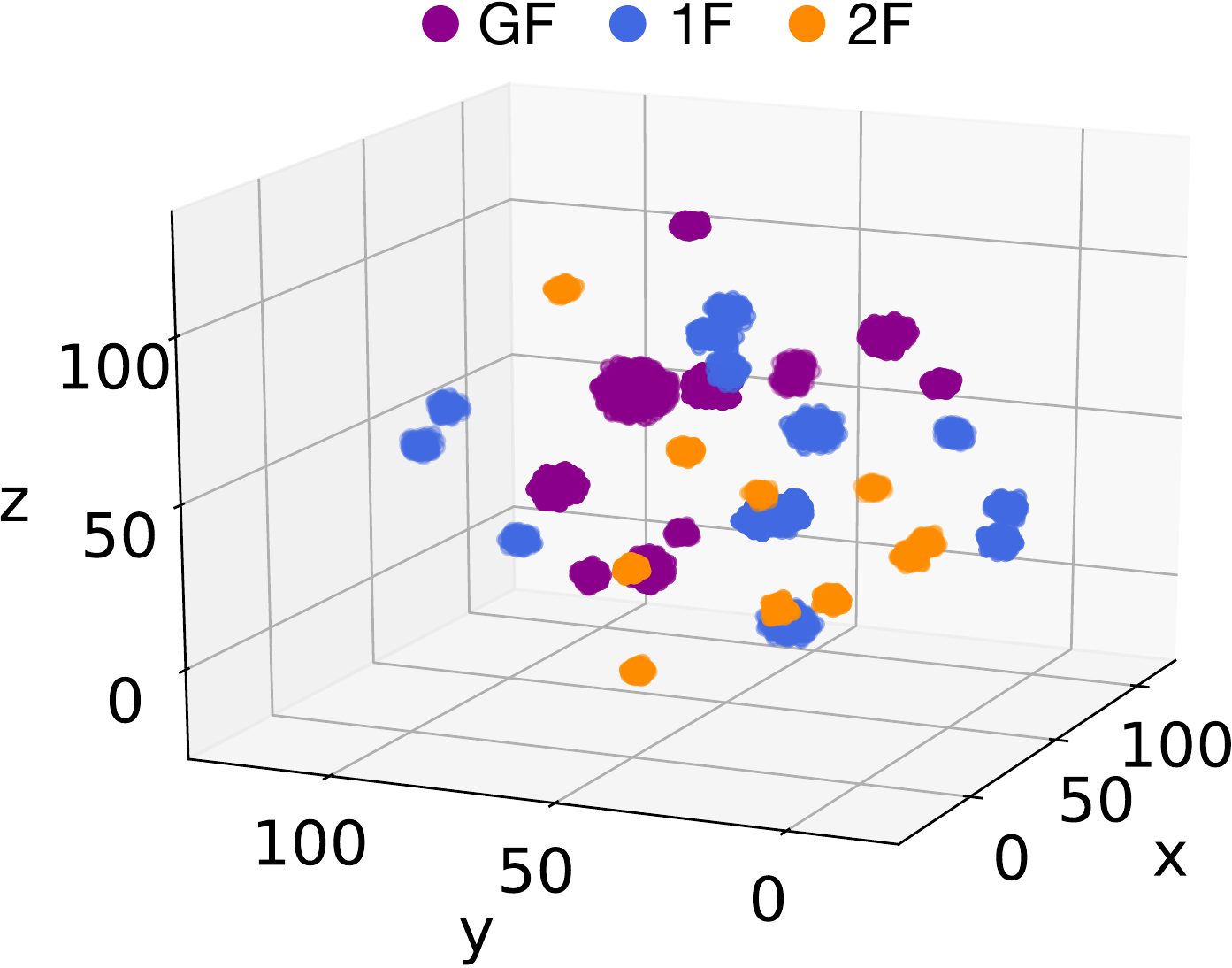}
        }
    \caption{The \linep{}'s embeddings of the RF signal samples from each floor naturally form a cluster in the embedding space, while MDS and autoencoder lead to poor embedding performance.}
    	\label{fig:embedding_floor}
    	\vspace{-0.2in}
\end{figure*}

Figure~\ref{fig:embedding_floor} visualizes how effective \linep{} maps RF signal samples collected in a three-story campus building into the (ego) embedding space, as compared to multidimensional scaling (MDS) and autoencoder used with a matrix representation of the RF signal samples. Here all the samples are `floor-labeled', and we use a visualization tool called t-SNE~\cite{van2008visualizing}. As shown in Figure~\ref{fig:embedding_floor}(a), \linep{} makes the embeddings of the signals collected on the same floor naturally form a cluster while separating different clusters apart. However, as can be seen from Figure~\ref{fig:embedding_floor}(b)--(c), MDS~\cite{cox2008multidimensional} and autoencoder~\cite{Goodfellow-et-al-2016} fail to cluster together the same-floor signals. Such unsatisfactory performance is expected to stem from the aforementioned missing value problem when the RF signal samples are represented in a matrix form.

\begin{figure}[t]
    \centering
    \includegraphics[width=0.35\textwidth]{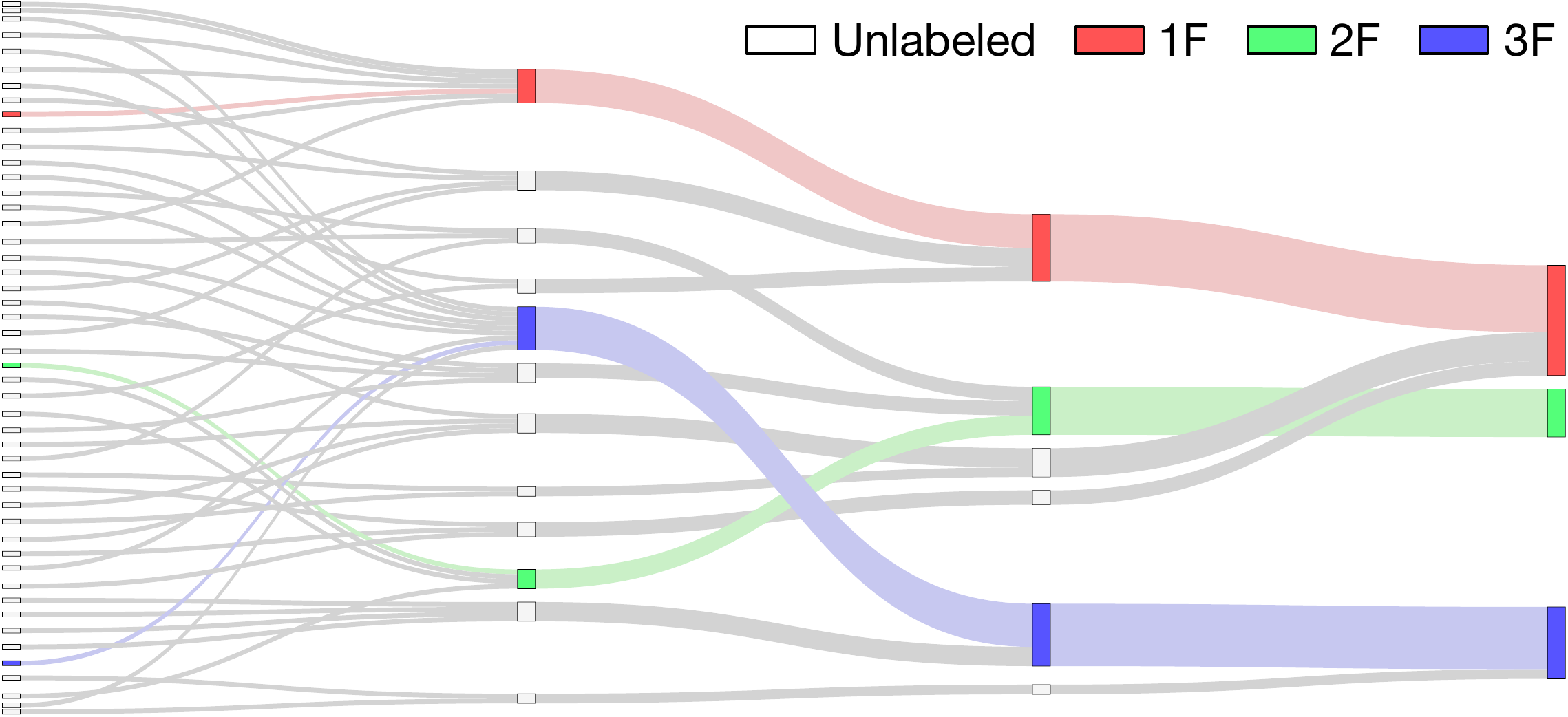}
    \caption{An illustrative example of proximity-based hierarchical clustering.}
    \label{fig:clustering}
    \vspace{-0.2in}
\end{figure}

\begin{figure*}[t]
    \centering
    \subfloat[20\%]{%
        \includegraphics[width=0.18\linewidth]{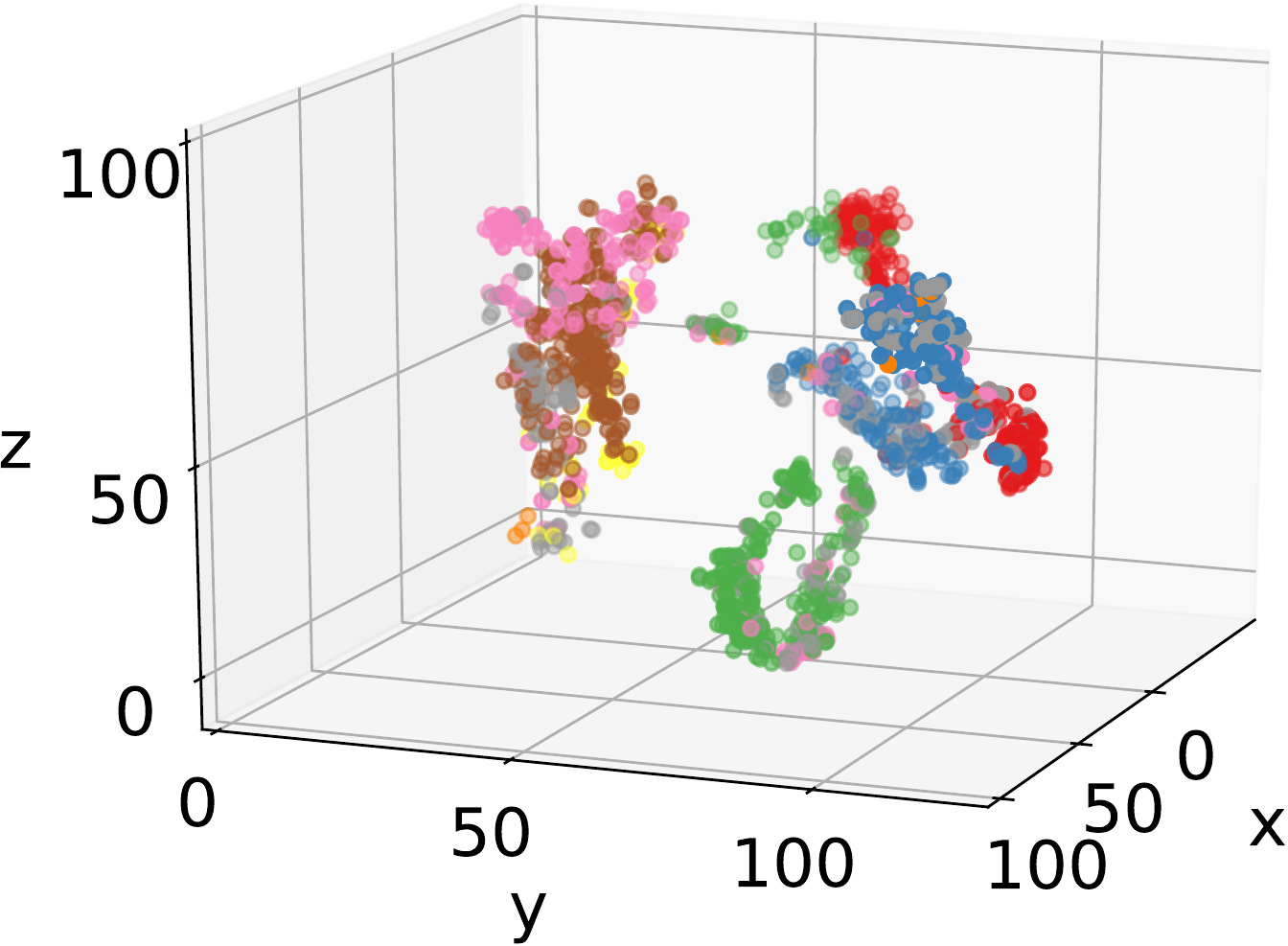}
        }
        \hspace{0.05in}
    \subfloat[40\%]{%
        \includegraphics[width=0.18\linewidth]{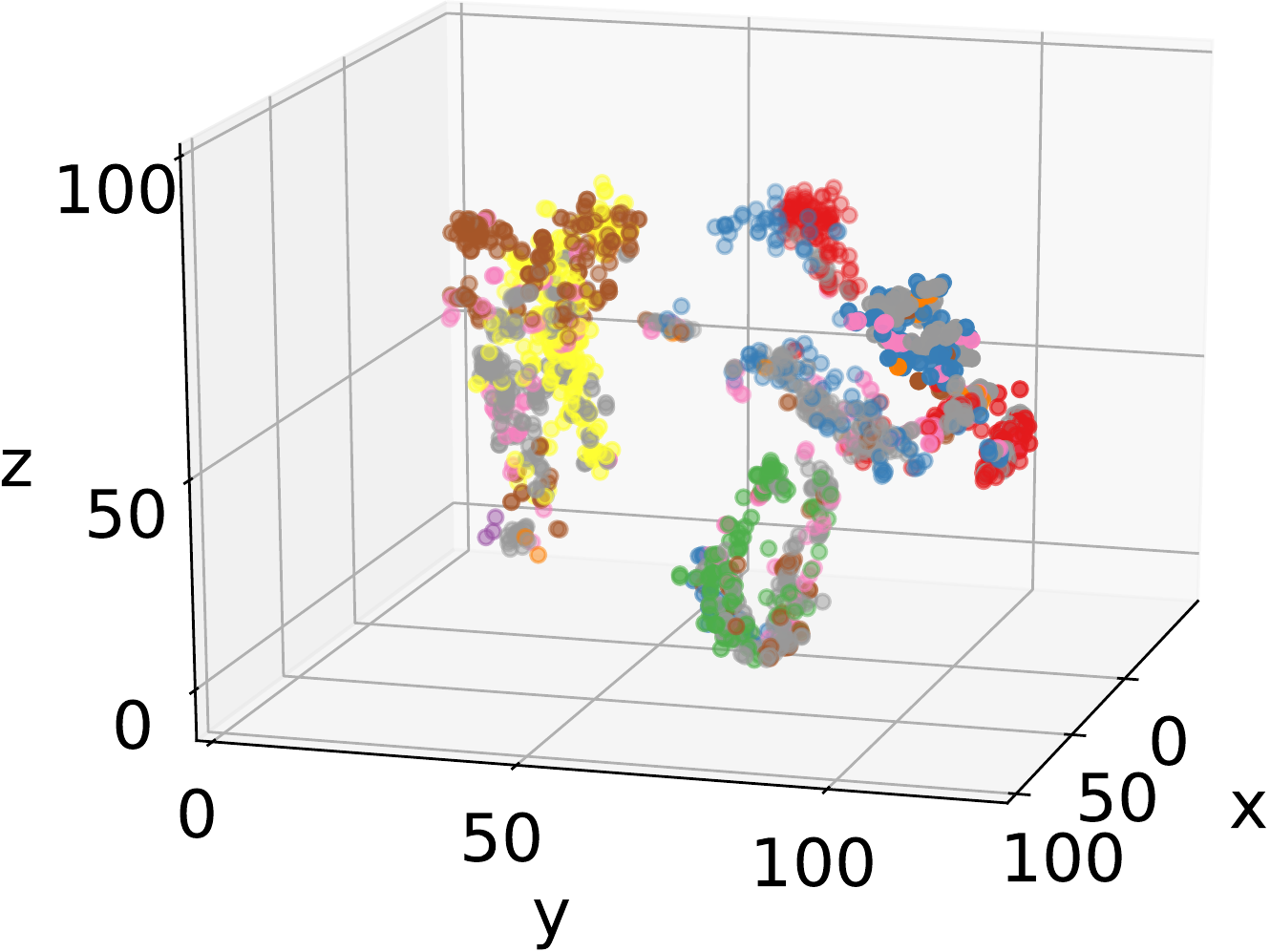}
        }
        \hspace{0.05in}
    \subfloat[60\%]{%
        \includegraphics[width=0.18\linewidth]{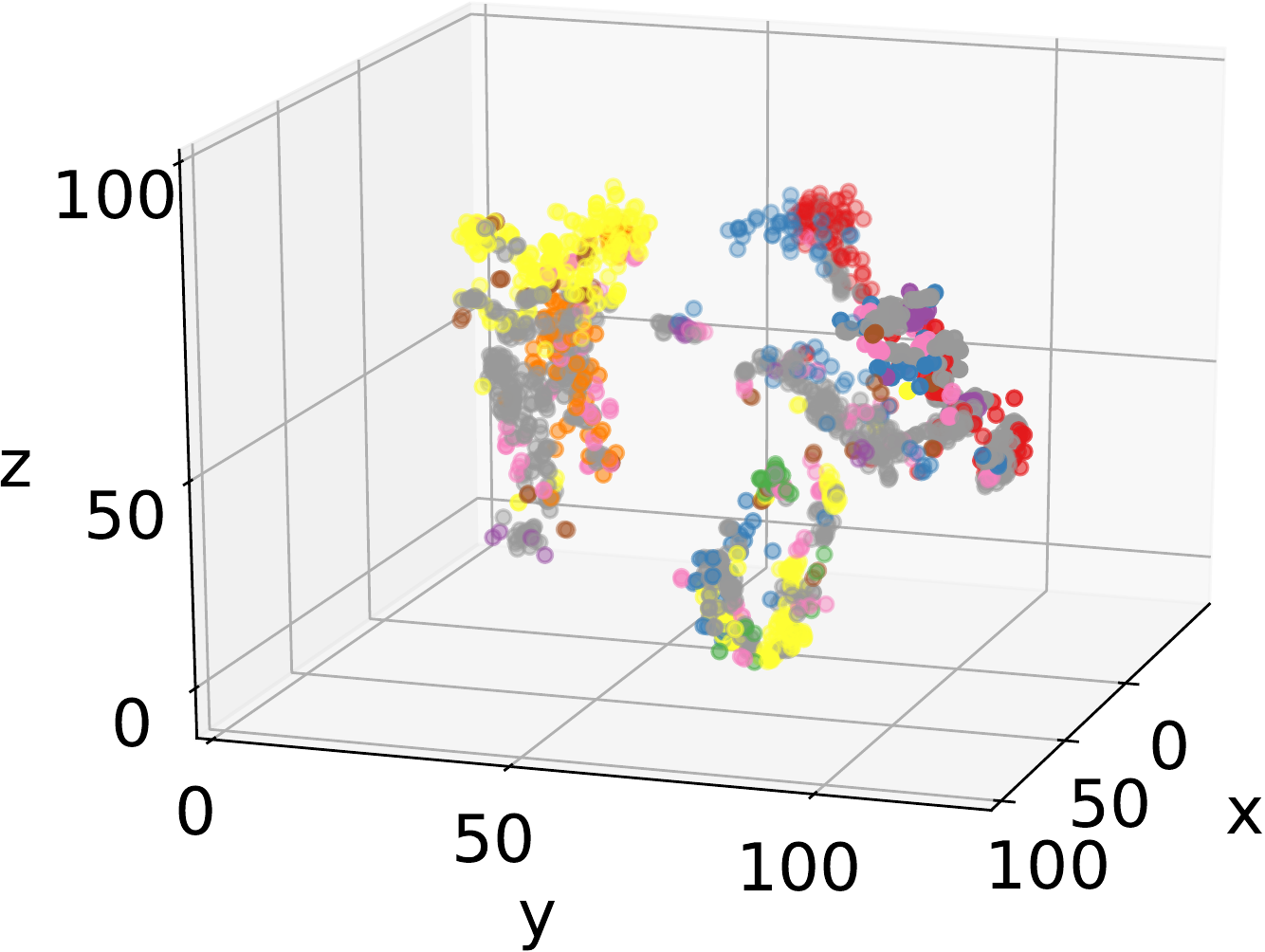}
        }
        \hspace{0.05in}
    \subfloat[80\%]{%
        \includegraphics[width=0.18\linewidth]{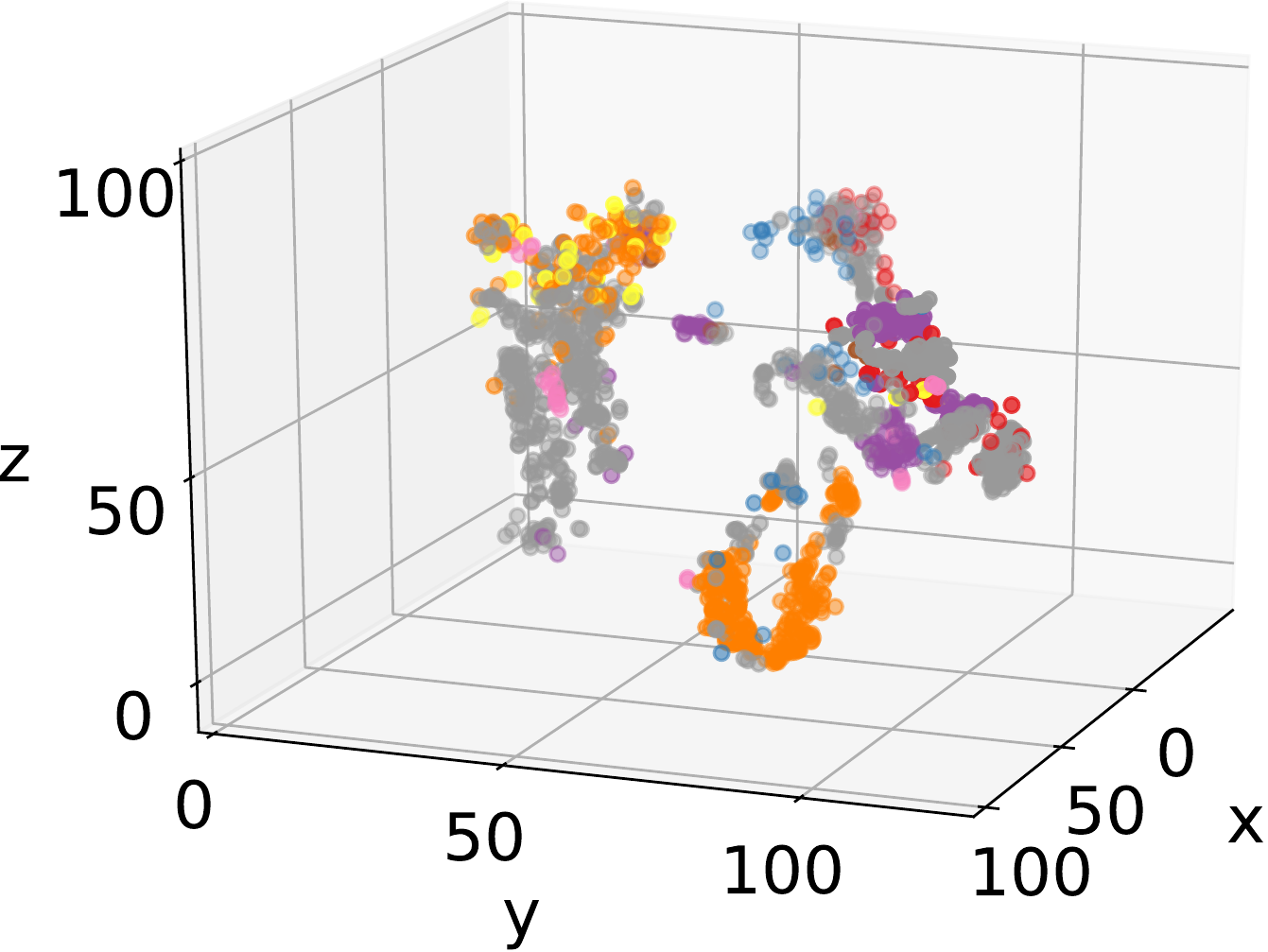}
        }
        \hspace{0.05in}
    \subfloat[100\%]{%
        \includegraphics[width=0.18\linewidth]{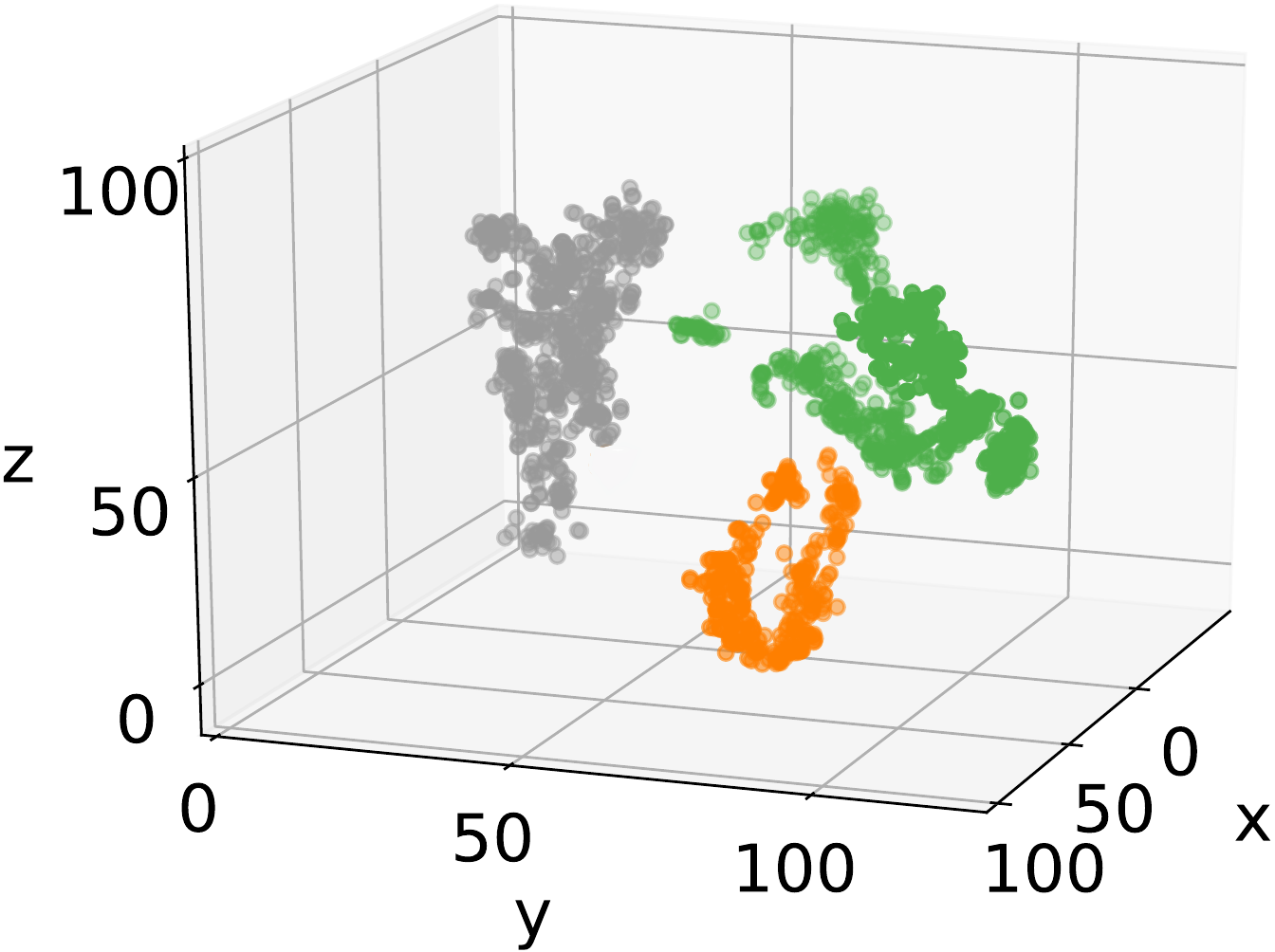}
        }
    \caption{\n{} gradually clusters the RF signal samples from the same floors in a three-story building, when only four samples are \emph{labeled} for each floor.}
    	\label{fig:clustering_process}
    	\vspace{-0.1in}
\end{figure*}

\subsection{Cluster Training}
\label{subsec:cluster_train}

Given the learned ego embeddings of RF signal samples, only a few of which come with floor labels (which floor they were collected), our goal here is to build a simple yet effective classification model. To this end, we use a proximity-based hierarchical clustering, which works as follows. Initially, each embedding is treated as an individual cluster. Note that only a few embeddings are for \emph{floor-labeled} samples, while most embeddings correspond to \emph{unlabeled} samples. We then repeatedly merge two clusters that are closest to each other such that the resulting cluster has the embedding of \emph{at most} one floor-labeled sample. In other words, two clusters that both have floor-labeled samples cannot be merged into a cluster. This clustering process is repeated until every embedding is merged into one cluster, and each cluster contains the embedding of exactly one floor-labeled sample, which is used as the \emph{floor label} of the cluster. Here, to compute the distance between two clusters, we use the following distance measure. For clusters $i$ and $j$, let $\bm{\Psi}_i$ and $\bm{\Psi}_j$ be the sets of the embeddings in their clusters, respectively. The distance between clusters $i$ and $j$ is defined as
\begin{equation}
d(\bm{\Psi}_i, \bm{\Psi}_j) := \frac{1}{|\bm{\Psi}_i| |\bm{\Psi}_j|}\sum_{\bm{u}_i \in \bm{\Psi}_i}\sum_{\bm{u}_j \in \bm{\Psi}_j} \| \bm{u}_{i}-\bm{u}_{j} \|_2,\label{eqn:dist_cluster}
\end{equation}
where $\|\cdot\|_2$ represents the $\ell_2$ norm.

Figure~\ref{fig:clustering} visualizes how the embeddings of 48 signal samples obtained on three floors are eventually merged into three clusters, each with the embedding of one floor-labeled sample. While this example assumes that there is only one labeled sample from each floor, the samples collected on each floor may possess several floor-labeled samples. Thus, multiple clusters can be associated with one floor. In addition, we provide a visualization of the clustering process in Figure~\ref{fig:clustering_process}, when there are four labeled samples each floor along with unlabeled samples obtained in the three-story building. As shown in Figure~\ref{fig:clustering_process}, all unlabeled samples are eventually merged into the clusters with labeled samples in the embedding space, which are well separated. Here the labeled samples are colored with the same color for each floor, and the unlabeled samples are colored with the color of the merged cluster.

\section{\n{}: Online Inference}
\label{sec:online}
We turn our attention to the online inference of \n{}. For a new RF signal sample, we explain how to obtain its ego and context embeddings. We then move on to the prediction of its floor label based on its ego embedding and the clusters obtained via the proximity-based hierarchical clustering.

\subsection{Embedding Prediction}
\label{subsec:network_predict}

When a new RF signal sample becomes available, it is added into the bipartite graph $\mathcal{G}$ as a new `RF-record' node, say $r$, and its edges are created with the `MAC' nodes that appear in the record. Some MAC nodes may also be newly added into the graph, if they are new ones. The edge weights are then determined based on the weight function in \eqref{eqn:edge_weight} and~\eqref{eqn:offset} with the recorded RSS values.\footnote{It is possible that the new signal sample only contains the MAC addresses that have \emph{never} been seen before. In this case, it may be the one collected outside the building and is thus discarded.}

Once the new node $r$ is added into the graph, its embeddings $\bm{u}_r$ and $\bm{u}'_r$ are learned to minimize the objective function in \eqref{eqn:objective} while the embeddings of the other nodes that have been in the graph remain fixed. This way, the new node $r$ is placed near its local neighborhood in the (ego) embedding space. Similarly for newly added MAC nodes. It is worth noting that minimizing the objective function in \eqref{eqn:objective} with respect to $\bm{u}_r$ and $\bm{u}'_r$ is computationally \emph{inexpensive} and can be done in real-time since the other (previously learned) embeddings remain fixed.

\subsection{Floor Prediction}
\label{subsec:floor_predict}

Given the ego embedding $\bm{u}_r$ of node/sample $r$, we calculate its distances with the centroids of all the clusters. We then find its predicted floor label as the label of the cluster that is closest to $\bm{u}_r$, which indicates on what floor the sample $r$ was collected. Specifically, the centroid $\bm{\psi}_i$ of cluster $i$ is defined as
\begin{equation*}
    \bm{\psi}_i := \frac{1}{|\bm{\Psi}_i|}\sum_{\bm{u}_i \in \bm{\Psi}_i} \bm{u}_i,
    \label{eqn:cluster_centroid}
\end{equation*}
and the distance between $\bm{u}_r$ and $\bm{\psi}_i$ is defined as
\begin{equation*}
    d(\bm{u}_r, \bm{\psi}_i) := \|\bm{u}_r - \bm{\psi}_i\|_2.
\end{equation*}
Let $l_i$ be the label of cluster $i$ and $l_{i^{\star}}$ be the label of the cluster whose centroid is closest to $\bm{u}_r$, where
\begin{equation*}
    i^{\star} := \arg\!\min_i d(\bm{u}_r, \bm{\psi}_i).
\end{equation*}
Then, the predicted label of sample $r$ is simply $l_{i^{\star}}$. Recall that each cluster has one floor-labeled sample, and the floor label of such a sample in the closest cluster becomes the predicted label of $r$. 

\section{Illustrative Experimental Results}
\label{sec:exp}
In this section, we present the extensive experiment results of \n{}. We first explain experiment settings and present the results on performance comparison between \n{} and state-of-the-art algorithms. We also study the impact of different system components and parameters on \n{}.

\begin{figure}[t]
    \centering
    \includegraphics[width=0.3\textwidth]{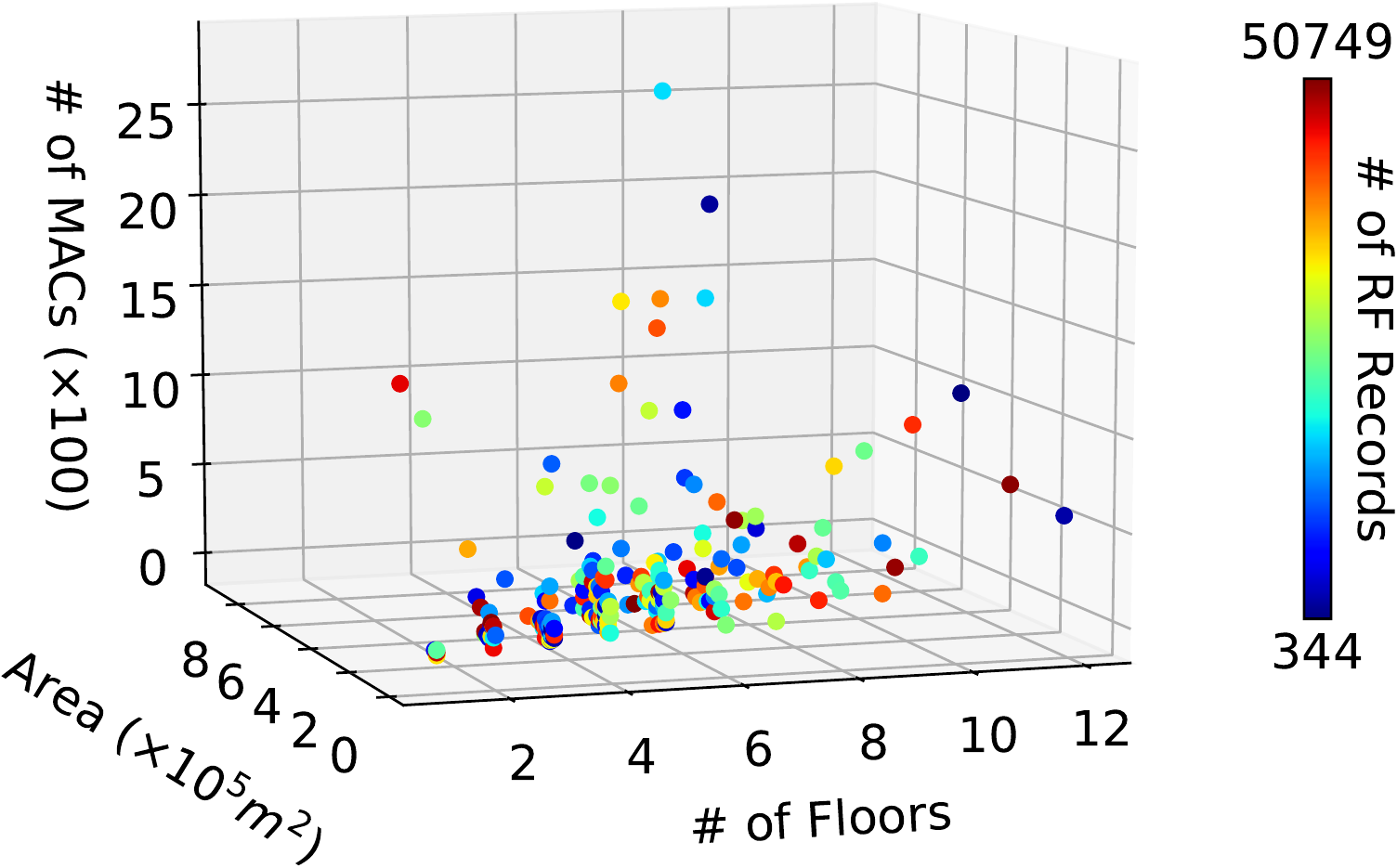}
    \vspace{-1mm}
    \caption{A summary of building information. Each point represents a building.}
    \label{fig:building_info}
    \vspace{-0.2in}
\end{figure}

\subsection{Experiment Settings}
\label{subsec:exp_settings}

\begin{figure}[t]
    \centering
    \includegraphics[width=0.28\textwidth]{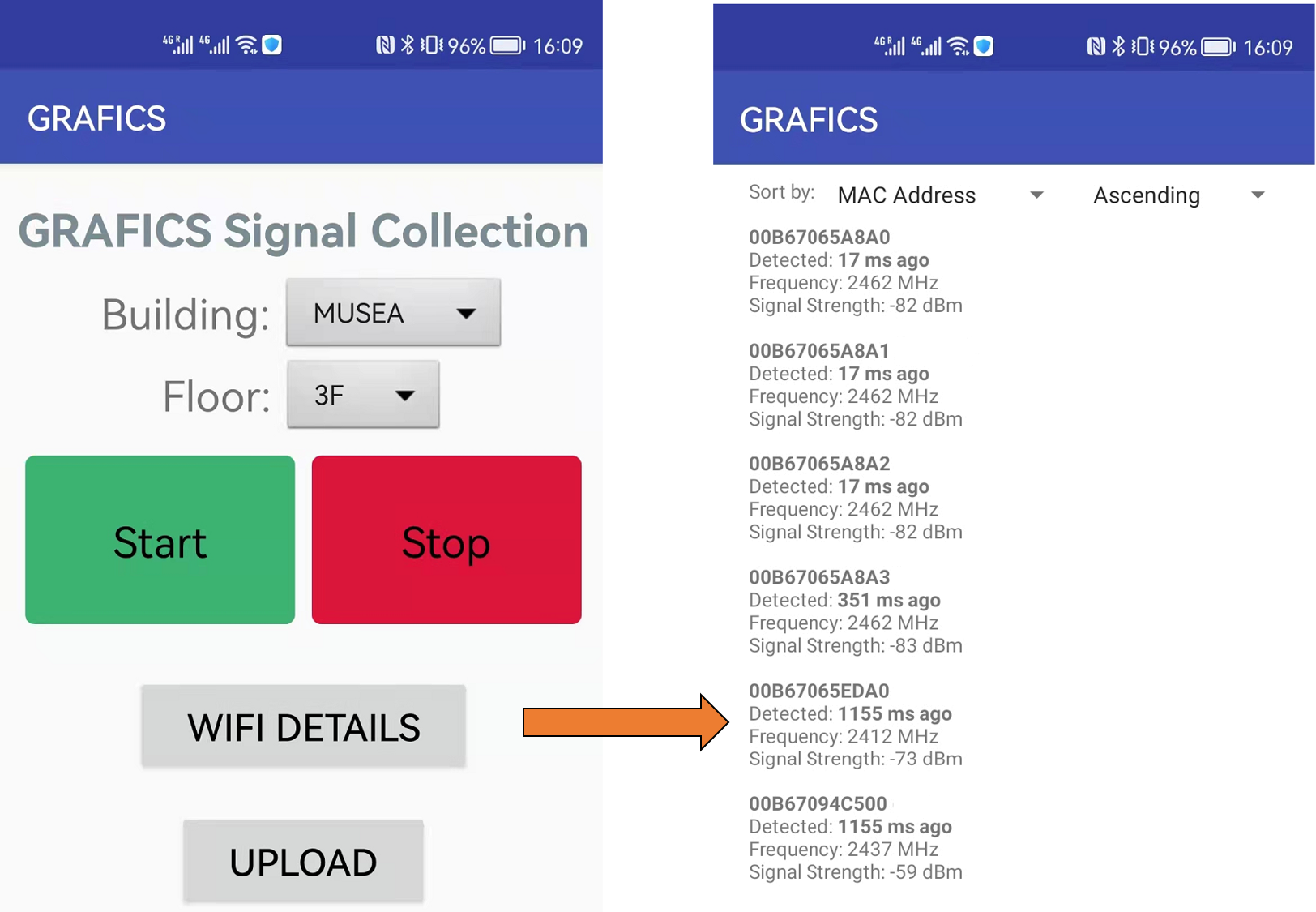}
    \caption{Screenshot of our data collection APP.}
    \vspace{-0.2in}
    \label{fig:app}
\end{figure}

\noindent\textbf{Experiment setup:} We conduct experiments on two large-scale datasets. The first one is Microsoft's open dataset in Kaggle competition~\cite{ms-kaggle}, which contains WiFi signal records collected in 204 buildings, where the smallest one is a two-story building and the tallest one has 12 floors. The other dataset is our own dataset. We develop and use a data collection APP (shown in Figure~\ref{fig:app}) to collect WiFi signal samples from five facilities in Hong Kong, which are two office towers, a hospital, and two shopping malls. As shown in Figure~\ref{fig:building_info}, the datasets cover a wide range of buildings in terms of the number of floors and the building size. The number of sensed MAC addresses varies over the buildings. The number of collected RF signal samples is also quite different across the buildings. Each \emph{floor} is associated with around 1000 WiFi signal samples on average, which are collected in a crowdsourced manner. Unless otherwise mentioned, we present the average results of the 204 buildings from the Microsoft dataset and the ones of the five buildings from the Hong Kong dataset separately.

We set the baseline parameters for performance comparison as follows. The learning and dropout rates of \linep{} are set to 0.001 and 0.1, respectively. We use eight-dimensional vectors for the ego and context embeddings in \linep{} and the embeddings for other algorithms. We use 70\% of each dataset for training and 30\% for testing. In the training samples, there are only four floor-labeled samples (which are randomly chosen) on each floor, while the rest of the samples are unlabeled. We run each algorithm 10 times for all test cases and report their average values. All experiments are conducted on the Ubuntu 18.04 server with 10 Intel Core I9-9900X cores at 3.5GHz, 64GB memory, and an NVidia 2080Ti graphic card. Our code is available online.\footnote{https://github.com/RobertFlame/Floors.}

\begin{figure*}[t]
\vspace{-1mm}
    \centering
    	\subfloat[Microsoft (log-scaled)]{%
        \includegraphics[width=0.24\linewidth]{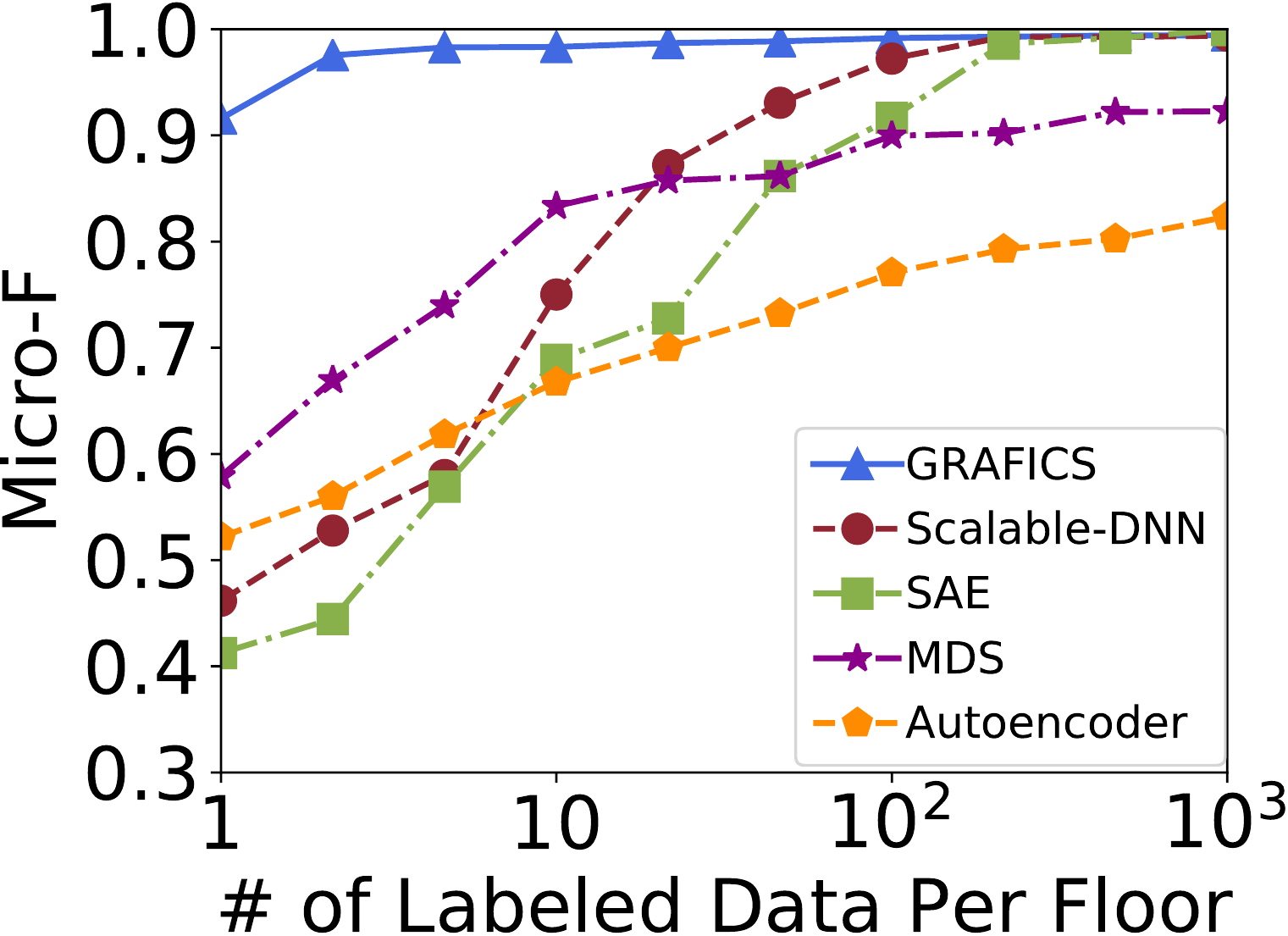}
        \includegraphics[width=0.24\linewidth]{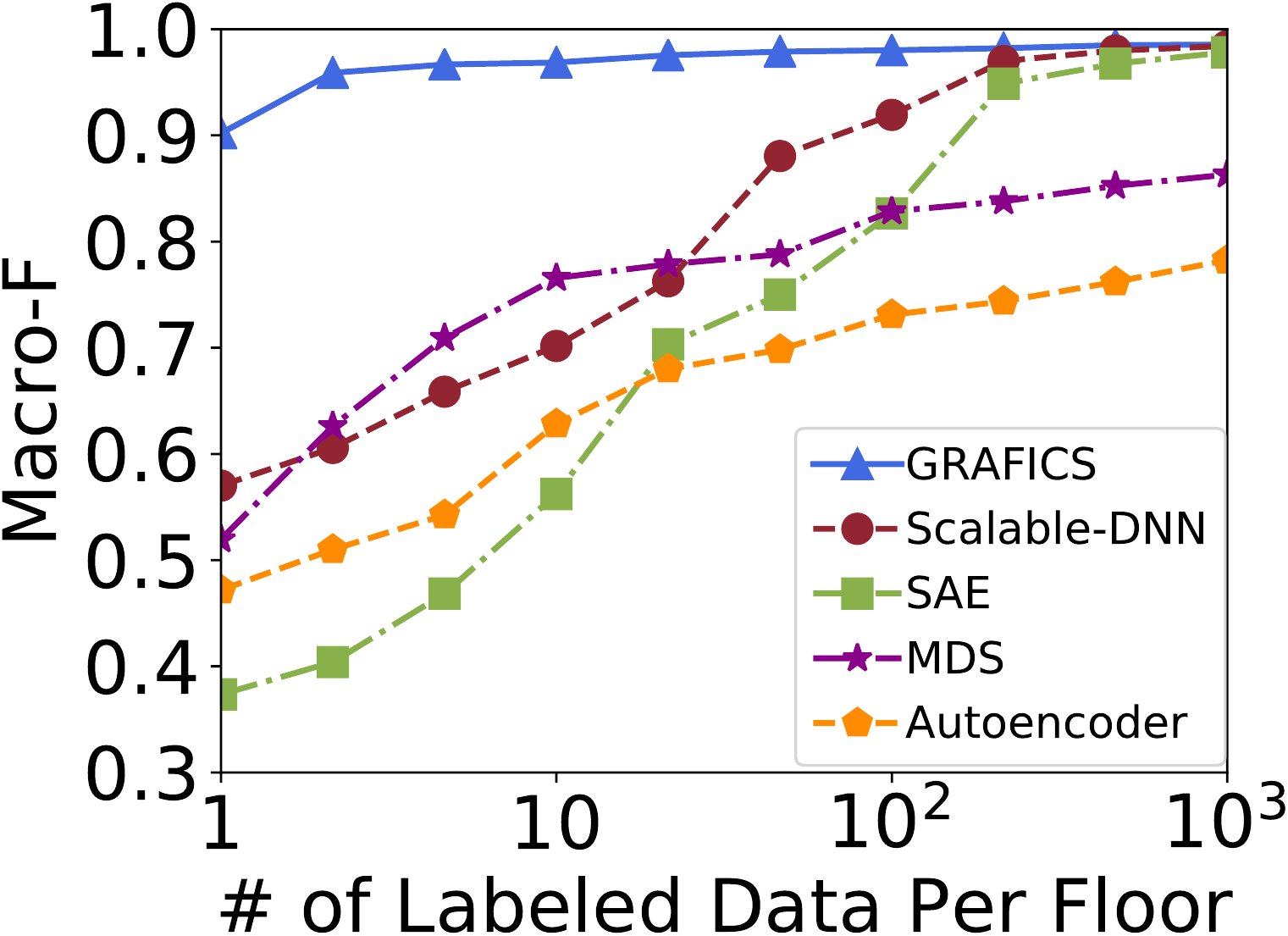}
        }
        \subfloat[Hong Kong (log-scaled)]{%
        \includegraphics[width=0.24\linewidth]{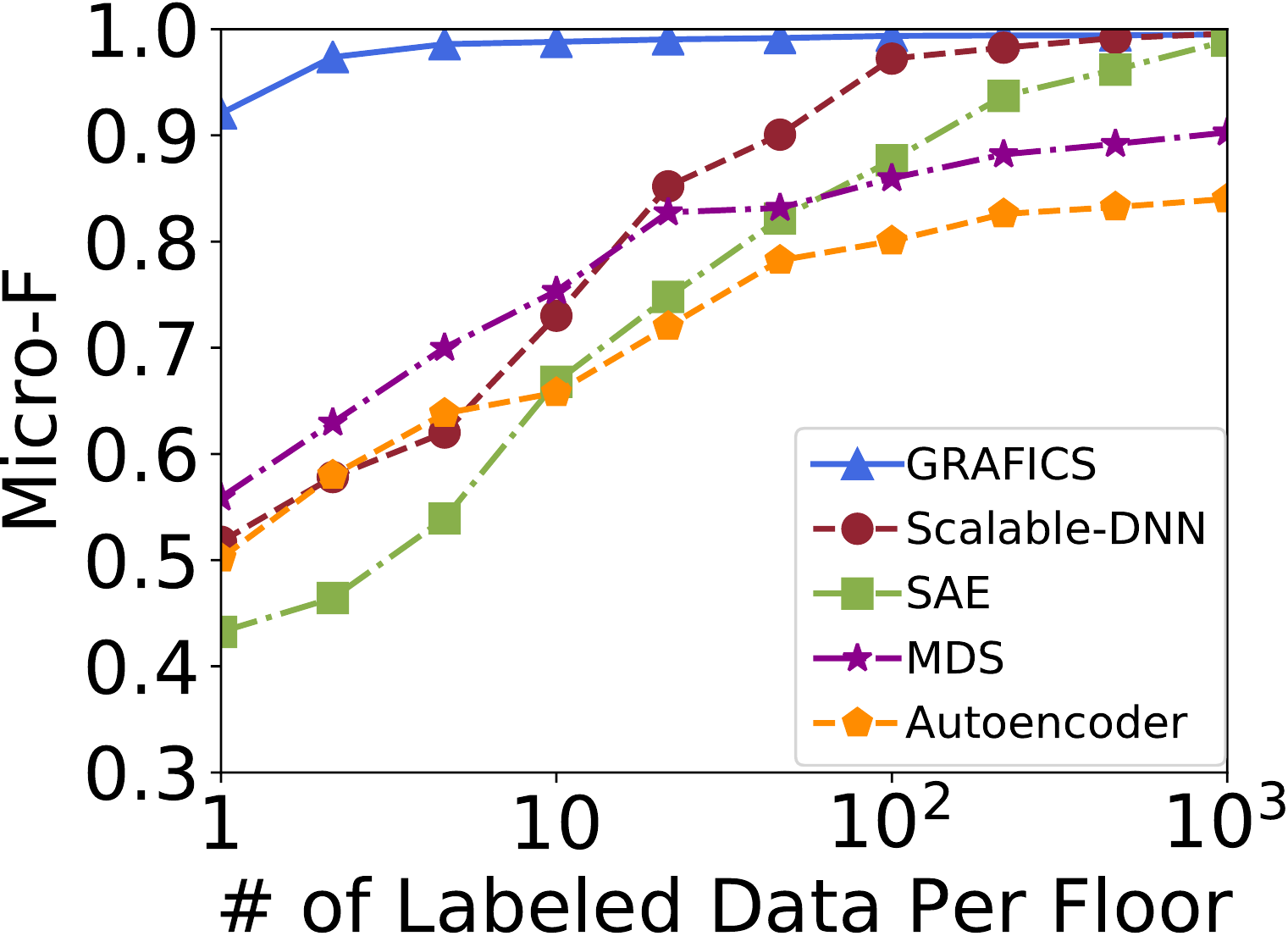}
        \includegraphics[width=0.24\linewidth]{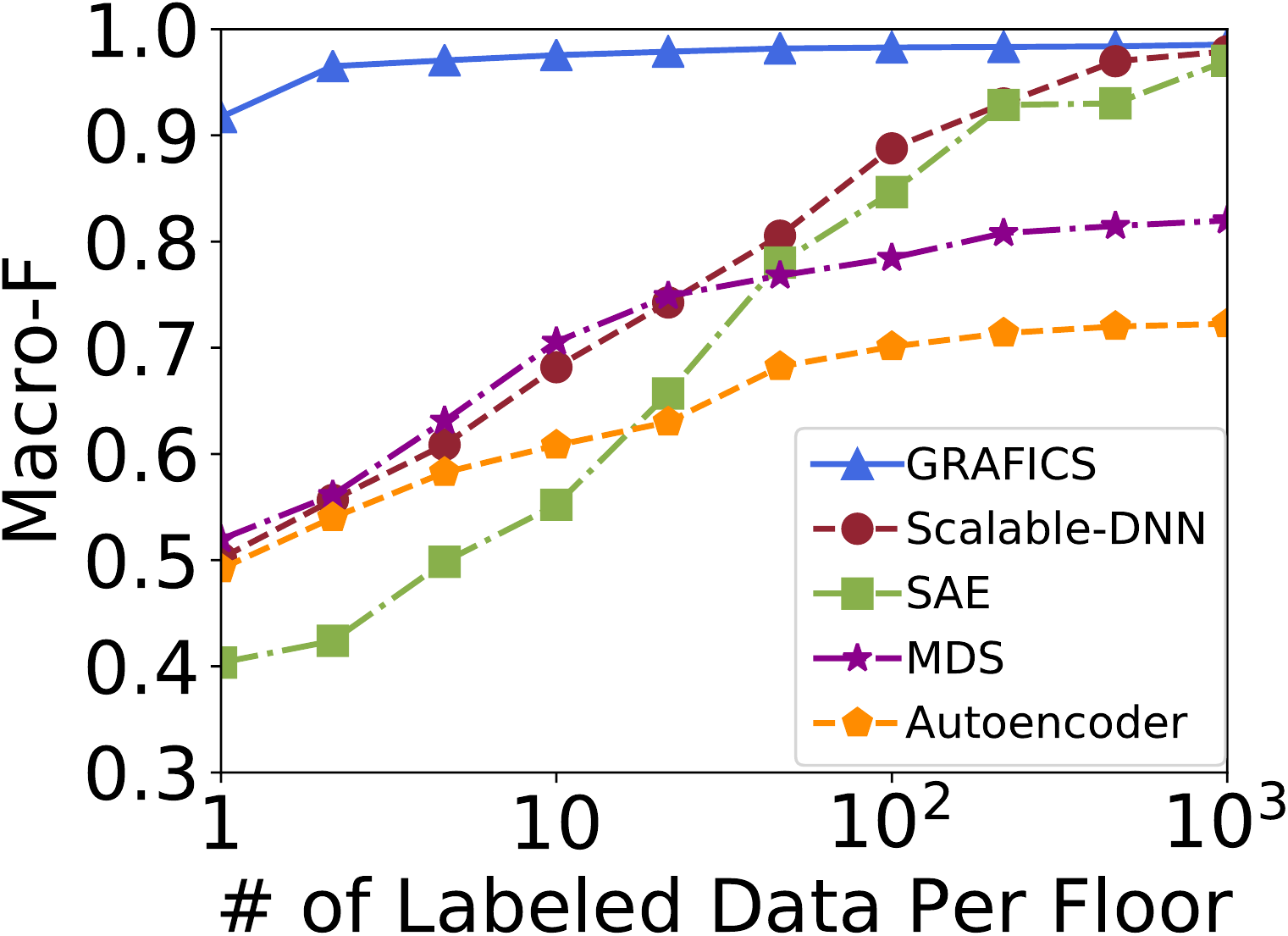}
        }
        \vspace{-1mm}
    	\caption{$F$-scores of \n{} with varying number of labeled samples on each floor.}
    	\vspace{-0.1in}
    	\label{fig:label_num_comparison}
\end{figure*}

\begin{figure*}[t]
    \centering
        \begin{spacing}{0.6}
        \includegraphics[width=0.55\linewidth]{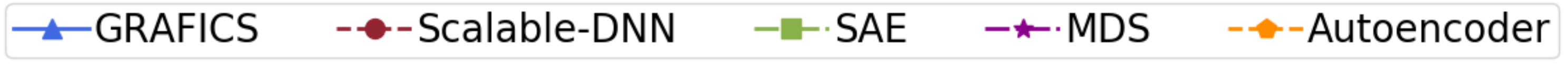}
        \end{spacing}
    	\subfloat[Microsoft (\#label = 4)]{%
        \includegraphics[width=0.24\linewidth]{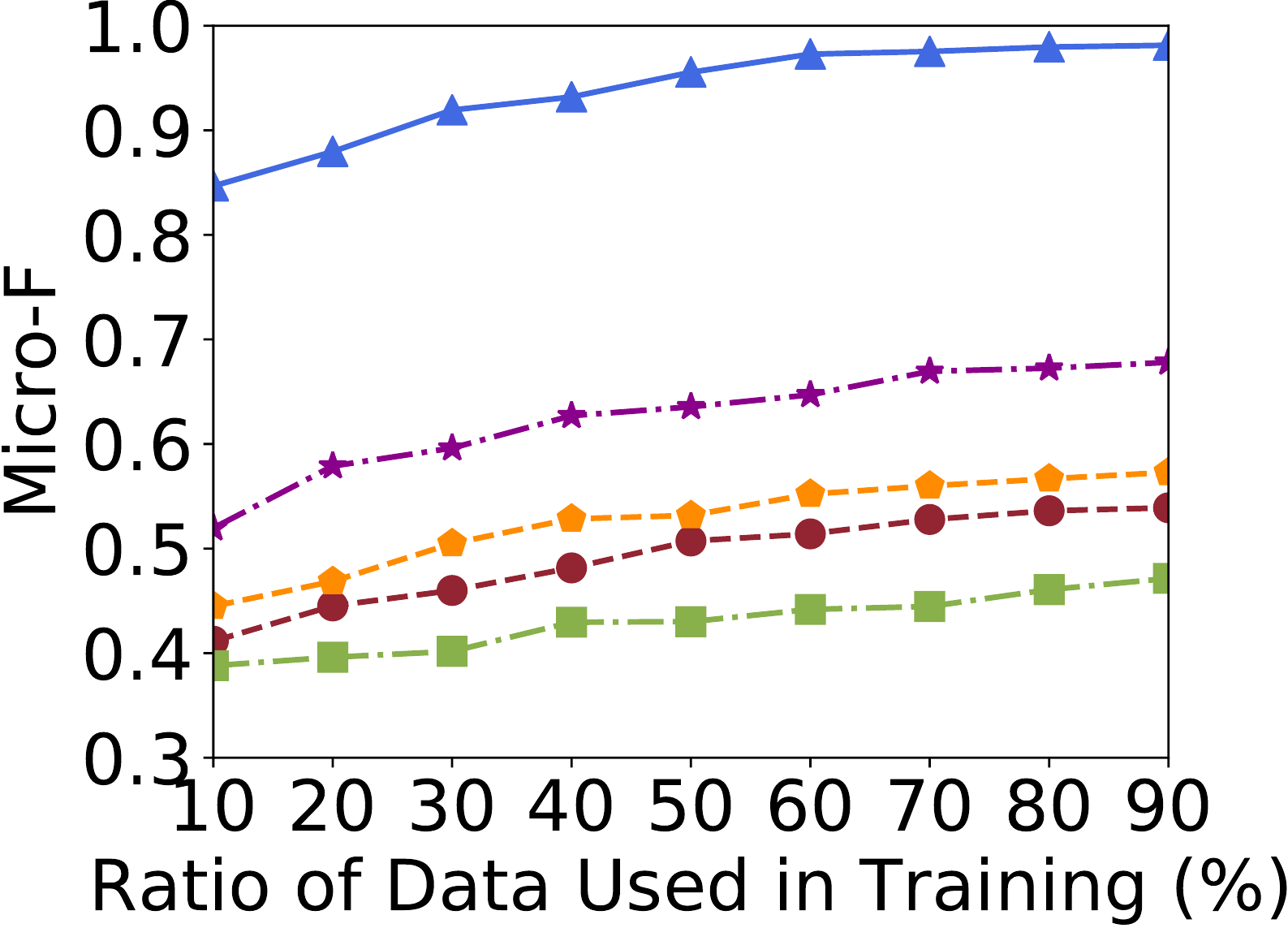}
        \includegraphics[width=0.24\linewidth]{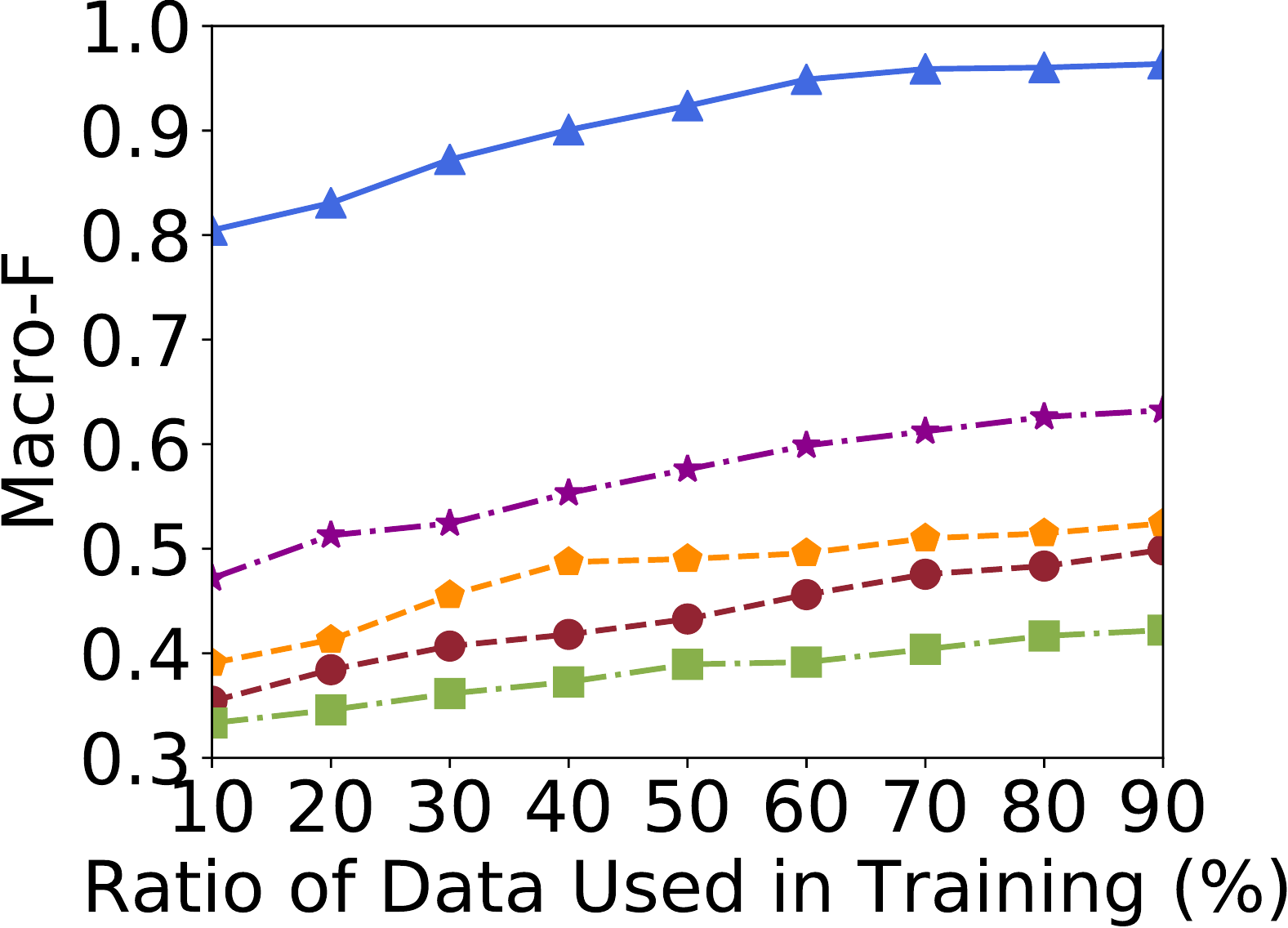}
        }
        \subfloat[Hong Kong (\#label = 4)]{%
        \includegraphics[width=0.24\linewidth]{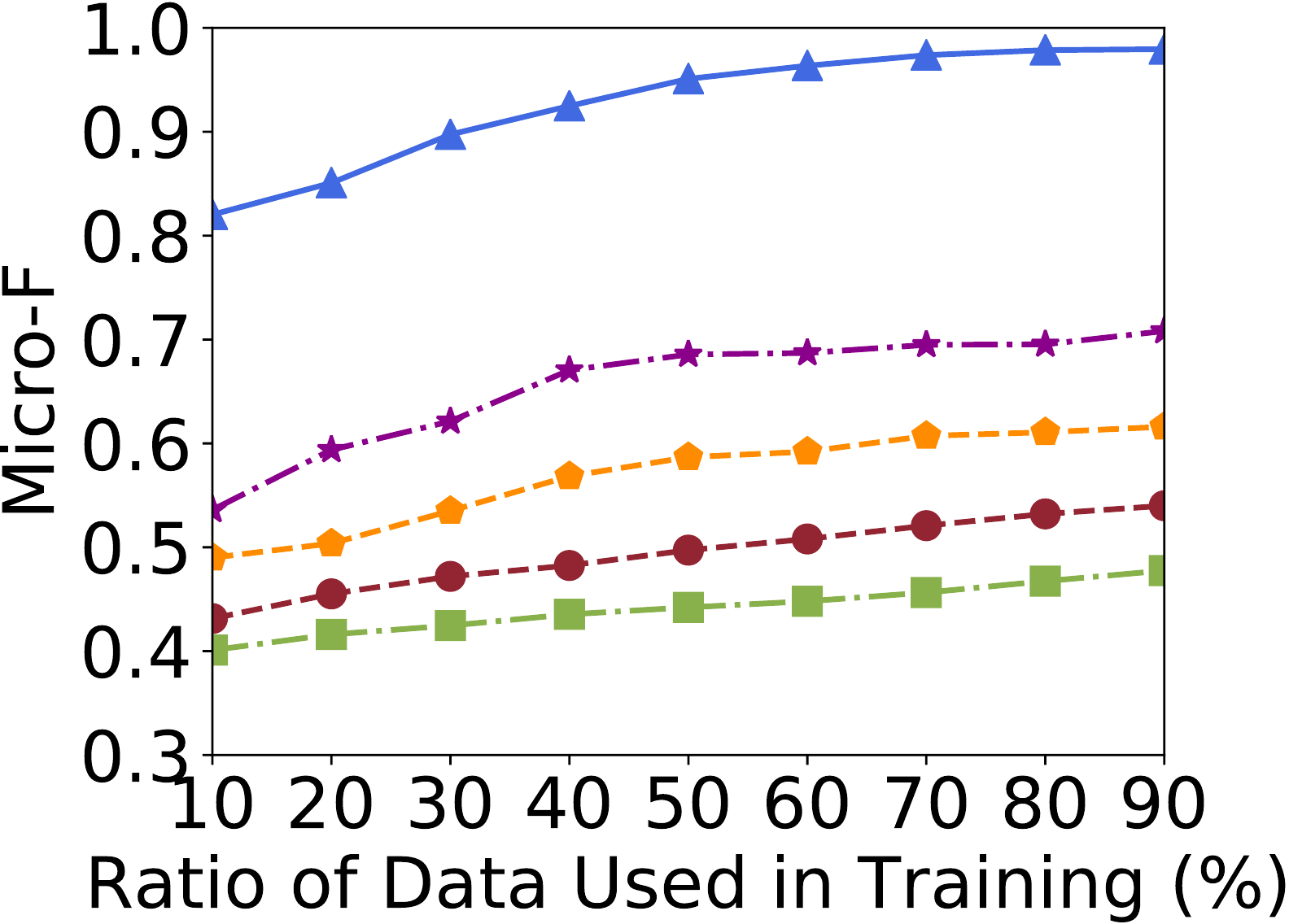}
        \includegraphics[width=0.24\linewidth]{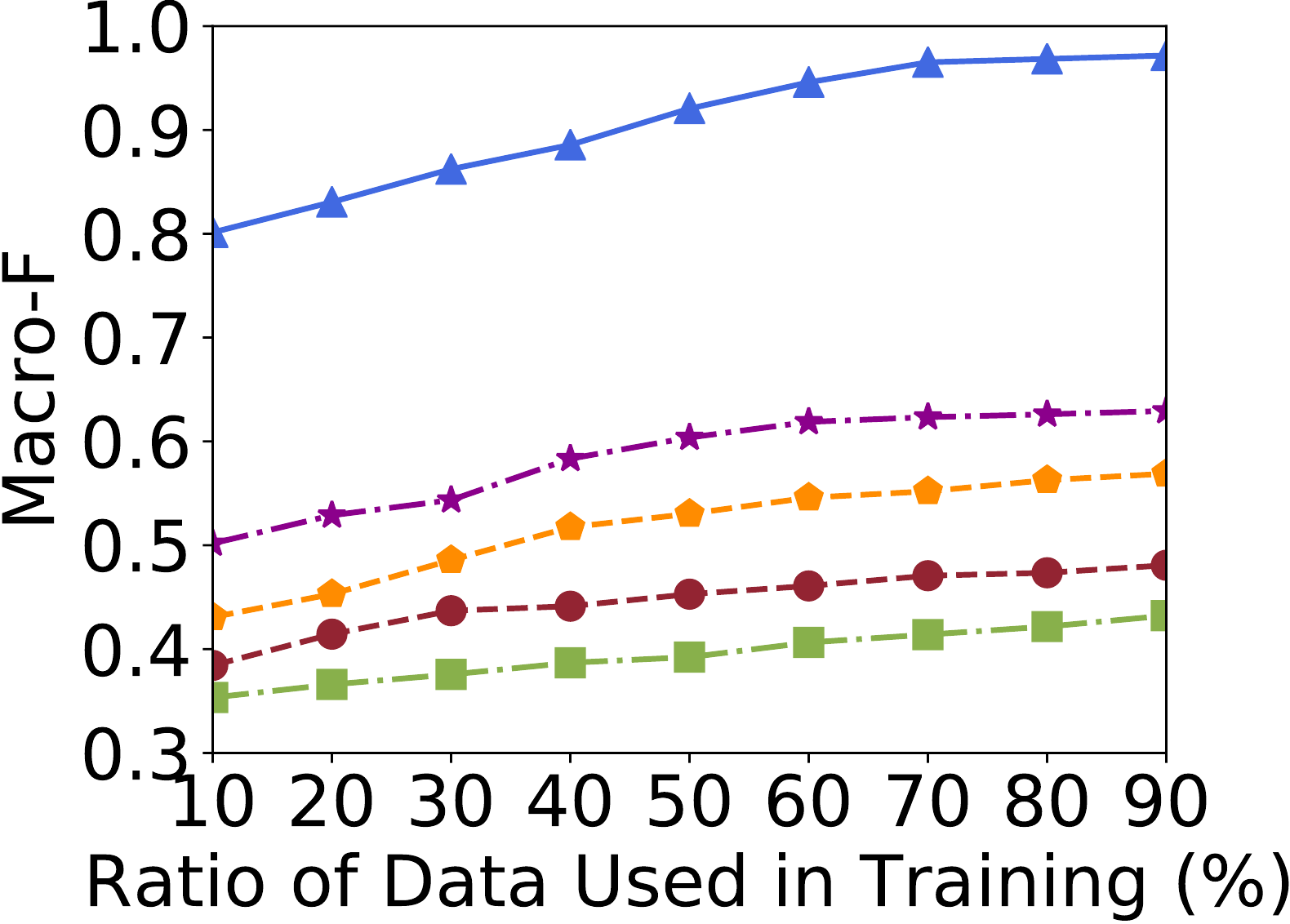}
        }
        \vspace{-1mm}
    	\caption{$F$-scores vs. the ratio of training data to the entire dataset, where the number of labeled samples is four on each floor.}
    	\vspace{-0.1in}
    	\label{fig:label_4}
\end{figure*}

\vspace{1mm}
\noindent\textbf{Algorithms used for comparison:} We consider the following state-of-the-art algorithms where two of them are combined with our proximity-based hierarchical clustering (denoted as \phc{}) for fair comparison:
\begin{itemize}[itemsep=2pt,leftmargin=1.5em]
    \item Scalable-DNN~\cite{kim2018scalable}: Embeddings are first generated through an encoding network, and floor ids are predicted as one-hot vectors through a feed-forward floor classifier.
    \item SAE~\cite{nowicki2017low}: Stacked autoencoders are used to learn low-dimensional embeddings, and a hierarchical classifier is then used for floor classification.
    \item Autoencoder~\cite{Goodfellow-et-al-2016} + \phc{}: It learns the embeddings of signal samples through an encoding and decoding process, which are then used with \phc{}.
    \item Multidimensional scaling (MDS)~\cite{cox2008multidimensional} + \phc{}: It learns the embeddings of samples by optimizing some distance matrix, which are then used with \phc{}.
\end{itemize}
For scalable-DNN and SAE, we set the parameters as described in~\cite{kim2018scalable} and~\cite{nowicki2017low}, respectively. Recall that a majority of training samples are unlabeled. We assign `pseudo' labels to the embeddings of unlabeled samples, which are the label of the closest labeled embedding, for training the supervised-learning models of scalable-DNN and SAE. The autoencoder consists of the four layers of 1-D convolution with the ReLU activation function. For MDS, the pairwise distance is set as 1 -- cosine similarity between two vectors.

\vspace{1mm}
\noindent\textbf{Evaluation metrics:} We use micro-$F$ and macro-$F$ to evaluate the classification performance. For floor $i$, we calculate the number of true positives $\text{TP}_i$, which is the number of correctly classified samples. We also calculate the number of false positives $\text{FP}_i$, i.e., the number of samples that are incorrectly labeled as $i$, and the number of false negatives $\text{FN}_i$, i.e., the number of samples misclassified as one of the other floors. Then, we have $P_i\!=\! \text{TP}_i/(\text{TP}_i\!+\!\text{FP}_i)$, $R_i = \text{TP}_i/(\text{TP}_i\!+\!\text{FN}_i)$ and $F_i = 2P_iR_i/(P_i\!+\!R_i)$. Assuming that there are $n$ floors in a building, we define the following micro metrics:
\begin{gather}
    \text{micro-}P=\frac{\sum_{i=1}^n \text{TP}_i}{\sum_{i=1}^n (\text{TP}_i \!+\! \text{FP}_i)},~ \text{micro-}R = \frac{\sum_{i=1}^n \text{TP}_i}{\sum_{i=1}^n (\text{TP}_i \!+\! \text{FN}_i)}, \nonumber \\
    \text{and}~ \text{micro-}F=2\frac{\text{micro-}P\times\text{micro-}R}{\text{micro-}P + \text{micro-}R}.
    \nonumber
\end{gather}
We also use the following macro metrics:
\begin{gather}
    \text{macro-}P=\frac{\sum_{i=1}^nP_i}{n},~ \text{macro-}R=\frac{\sum_{i=1}^nR_i}{n},~ \text{and}
    \nonumber \\
    \text{macro-}F=2\frac{\text{macro-}P\times\text{macro-}R}{\text{macro-}P + \text{macro-}R}.
    \nonumber
\end{gather}

\subsection{Comparison with State-of-the-art Algorithms}
\label{subsec:eval_overall}

In Figure~\ref{fig:label_num_comparison}, we provide comparison results of \n{} against other state-of-the-art algorithms for floor classification with varying number of labeled samples on each floor. As can be seen from Figure~\ref{fig:label_num_comparison}, more labeled samples are available, better the performance of each algorithm. Nonetheless, \n{} achieves the best overall performance. It is even able to achieve very high prediction accuracy with just a few labeled samples (i.e., four samples) on each floor, thanks to its novel bipartite graph modeling, high-quality graph embedding via \linep{}, and effective hierarchical clustering. 

Scalable-DNN and SAE, however, require a large number of labeled samples. This is somewhat well expected due to their supervised learning nature where their models need to be calibrated with abundant labeled samples. To reach comparable $F$-scores to the ones with \n{}, they need about $400$ to $700$ labeled samples per floor, which are $100\times$ more than the ones required by \n{} and would be difficult to obtain in practice. On the other hand, MDS and autoencoder do not have much benefit from having more labeled samples since their embeddings are not as good as the ones by  \linep{} due to the missing value problem.

We are also interested in how the ratio of training data to the entire dataset affects the performance of each model while the number of labeled samples for training remains unchanged. The results are shown in Figure~\ref{fig:label_4}, where the numbers of labeled samples remain fixed as four. We observe that the performance of every model improves with increasing number of samples for training.

\subsection{System Component Study}
\label{subsec:sys_component}

\begin{figure*}[t]
    \centering
        \subfloat[Microsoft (\#label = 4)]{%
        \includegraphics[width=0.23\textwidth]{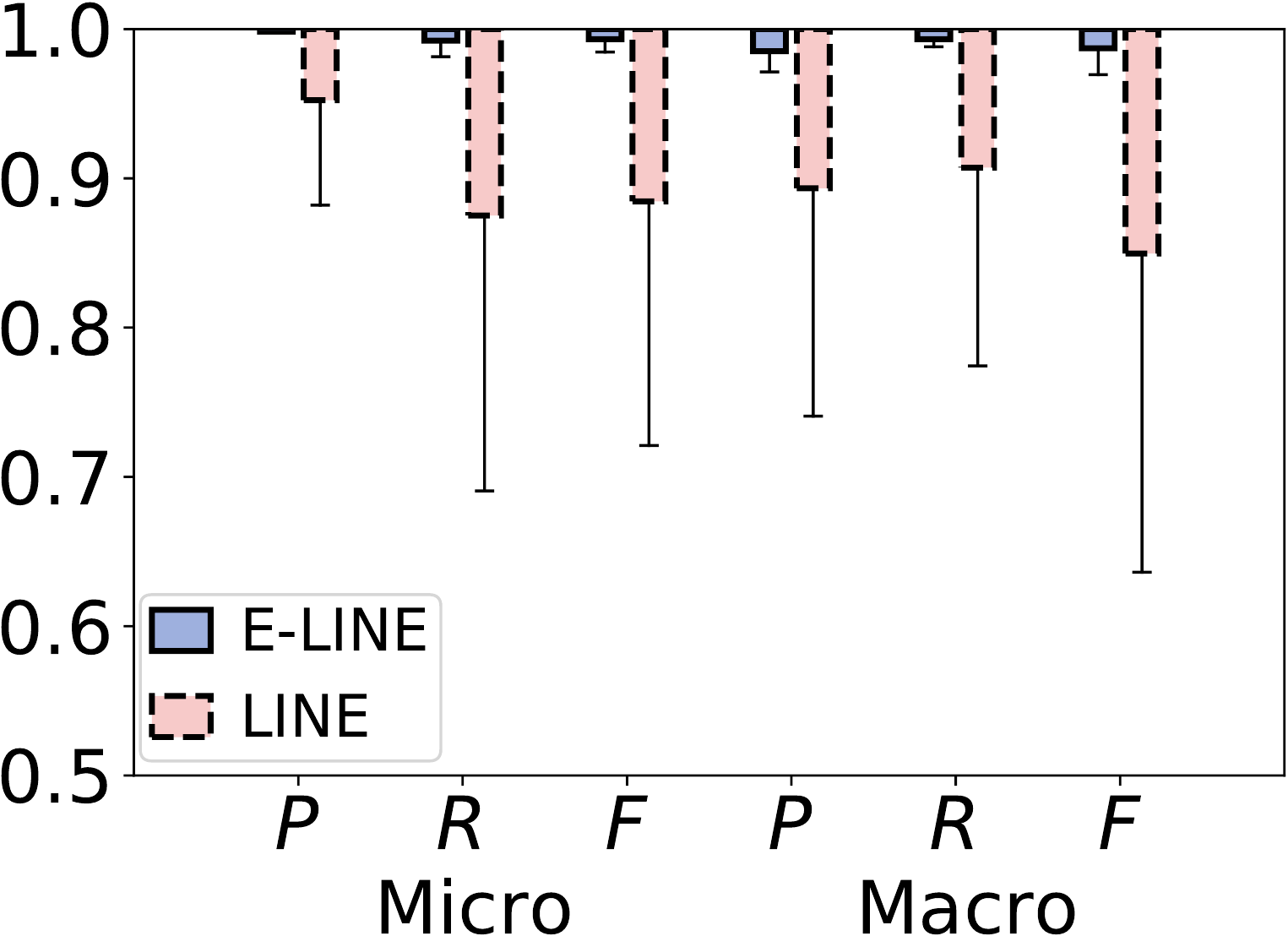}
        }
        \hspace{0.04in}
        \subfloat[Hong Kong (\#label = 4)]{%
        \includegraphics[width=0.23\textwidth]{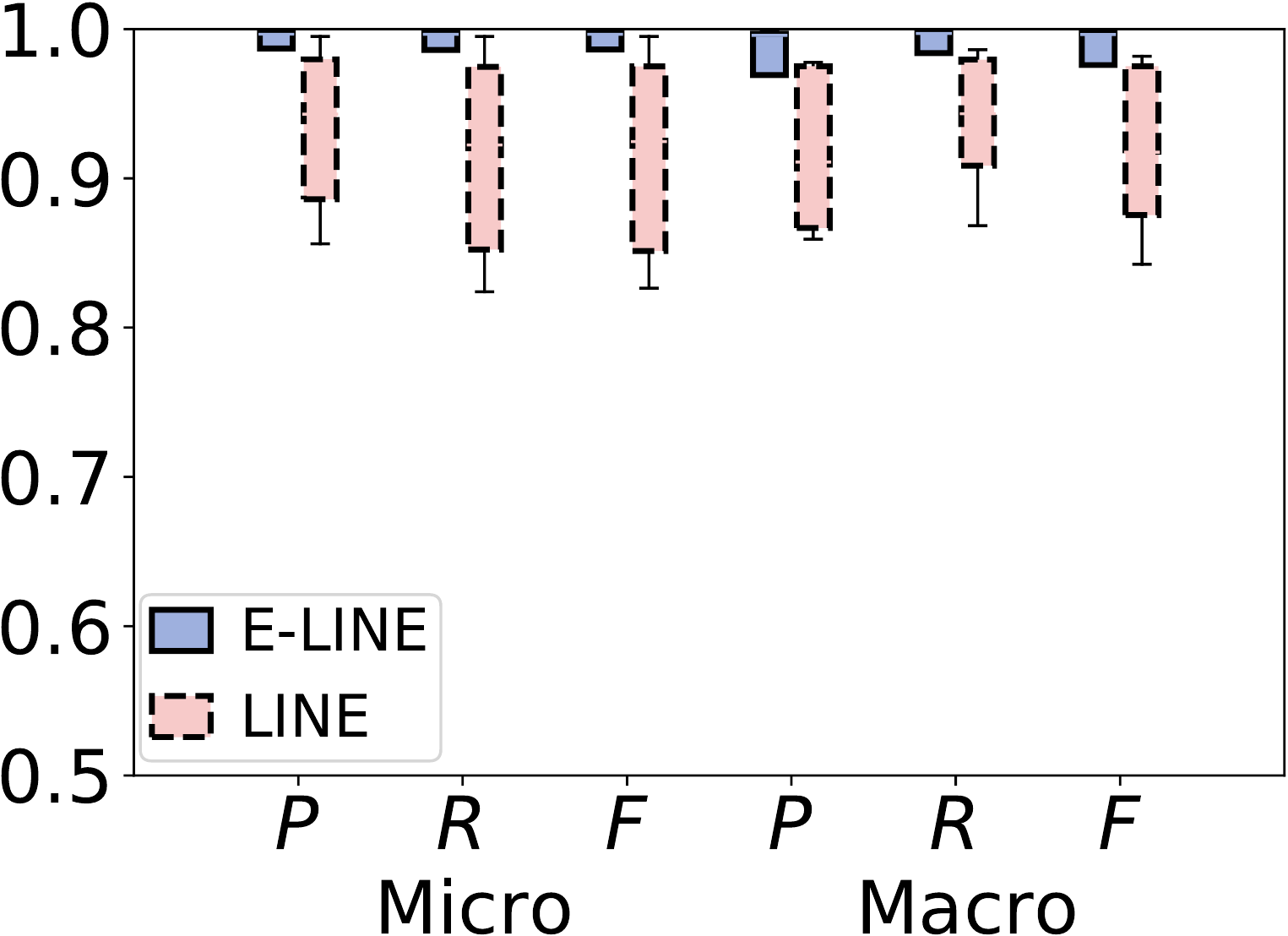}
        }
        \hspace{0.04in}
    	\subfloat[Microsoft (\#label = 40)]{%
        \includegraphics[width=0.23\textwidth]{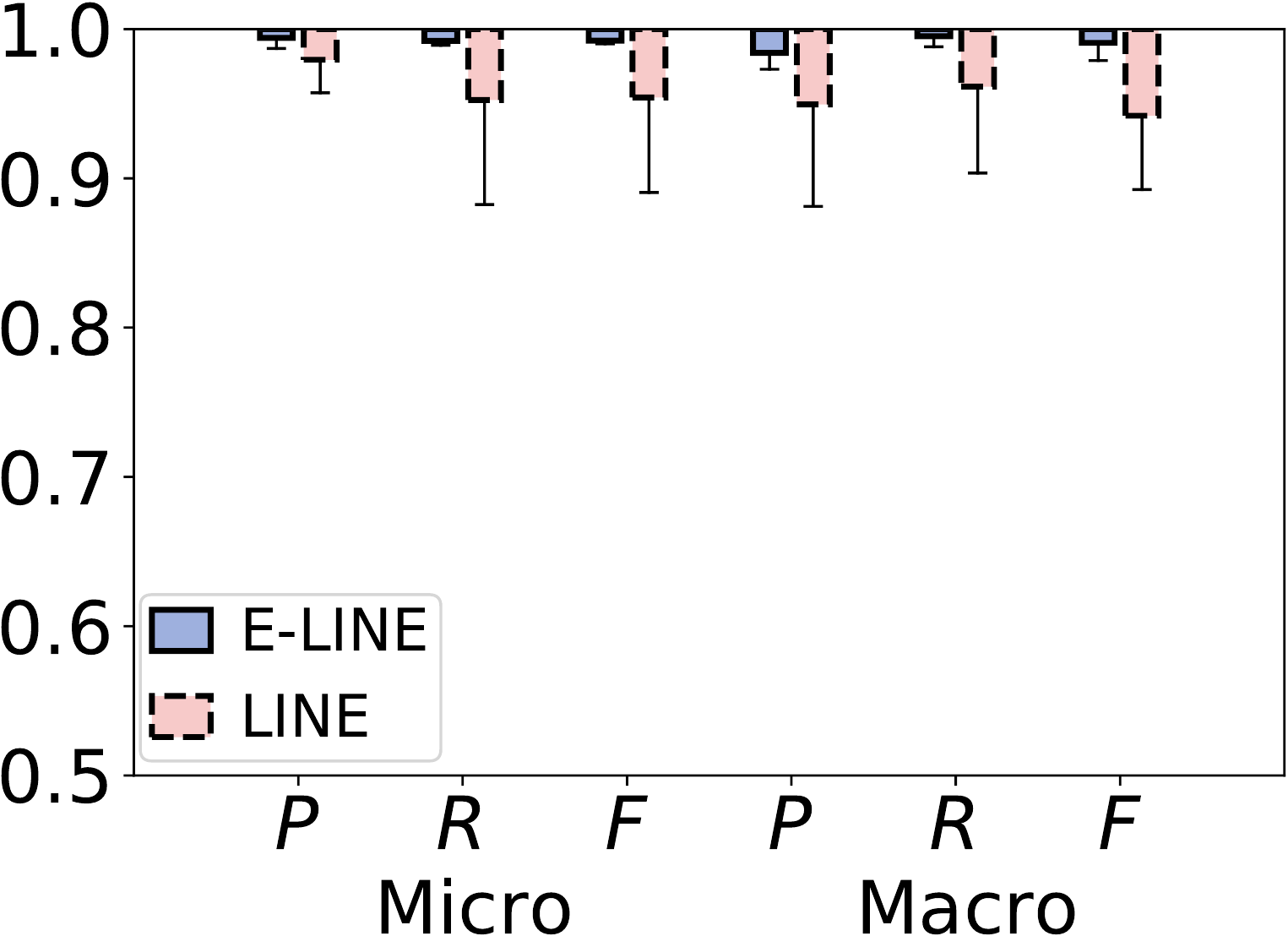}
        }
        \hspace{0.04in}
        \subfloat[Hong Kong (\#label = 40)]{%
        \includegraphics[width=0.23\textwidth]{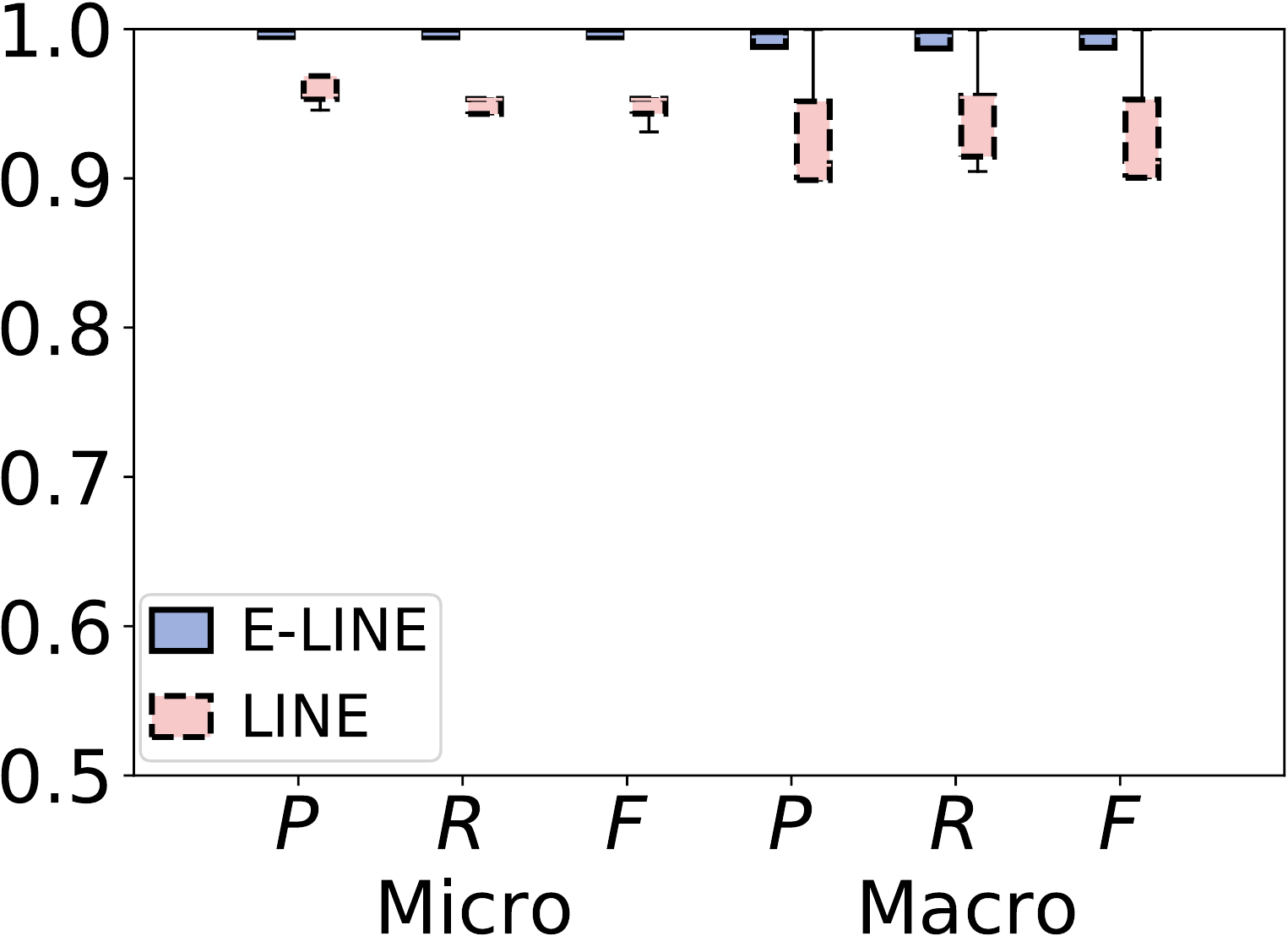}
        }
    	\caption{Performance comparison between \n{} (with \linep{}) and \n{} with LINE.}
    	\label{fig:box_eline}
    	\vspace{-0.4in}
\end{figure*}

To see how much \n{} benefits from \linep{} over LINE, we present the comparison results between \n{} and \n{} with LINE (instead of \linep{}) in Figure~\ref{fig:box_eline}. Here we consider two cases where one is with four labeled samples per floor and the other is with 40 labeled samples per floor. For LINE, we consider its second-order proximity only since it turns out to be better than LINE with first-order and second-order proximities. We omit the results due to space limit. As can be seen from Figure~\ref{fig:box_eline}, when four labeled samples per floor are only available, \n{} with LINE does not perform well and it exhibits high variance in its performance. As the number of available labeled samples increases, the performance of \n{} with LINE greatly improves and becomes more stable. However, \n{} already achieves the ideal prediction accuracy only with four labeled samples. The results confirm the advantage of \linep{} that learns the embeddings of nodes based on their local neighborhood information, which is more than just their one-hop neighborhood information -- the case with LINE. In other words, \linep{} makes the embeddings of nodes in their local neighborhood similar to each other while letting the ones of distant nodes dissimilar to each other.

\begin{figure}[t]
    \centering
    	\subfloat[Microsoft]{%
        \includegraphics[width=0.47\linewidth]{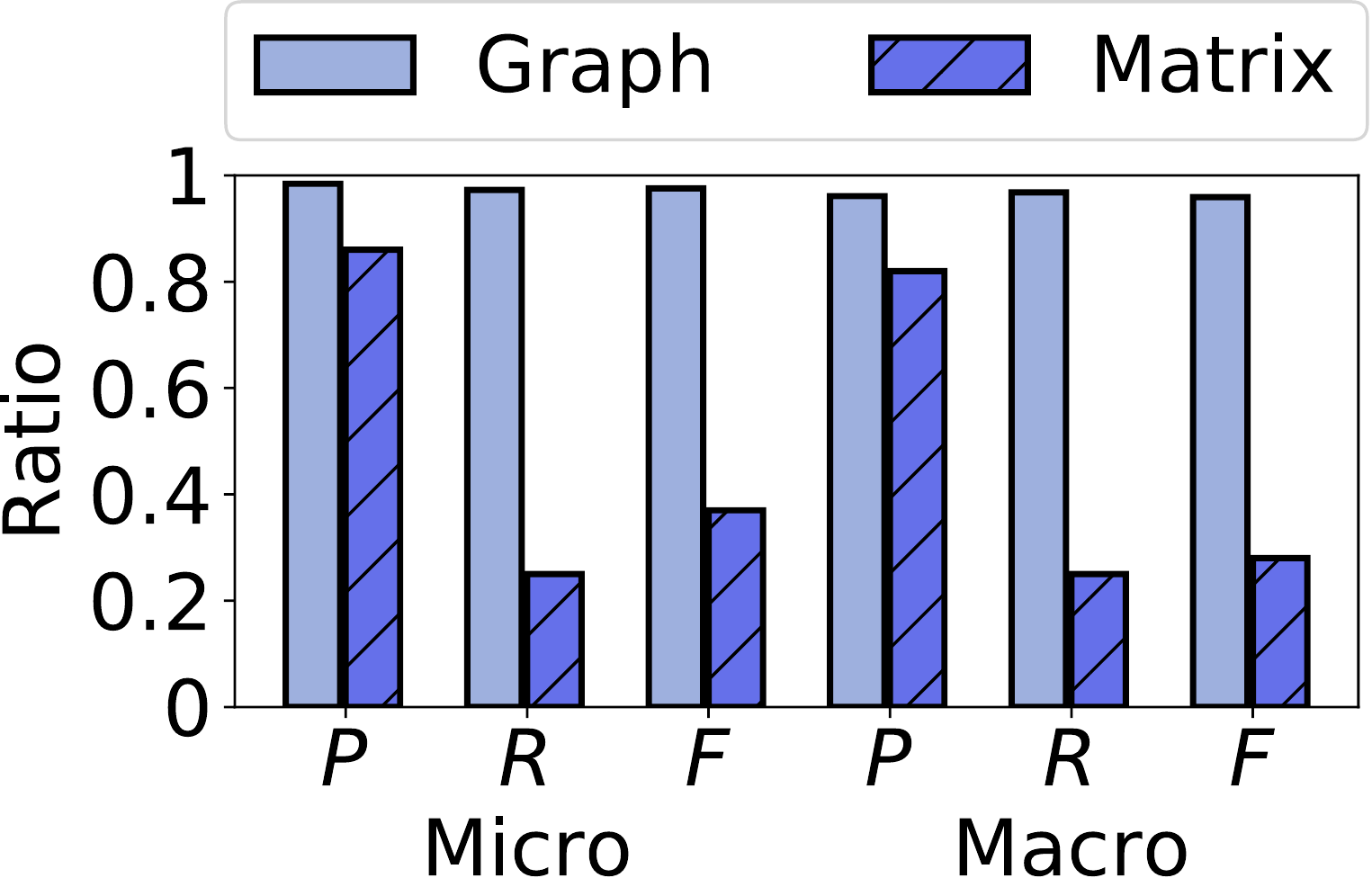}
        }
        \hspace{0.02in}
        \subfloat[Hong Kong]{%
        \includegraphics[width=0.47\linewidth]{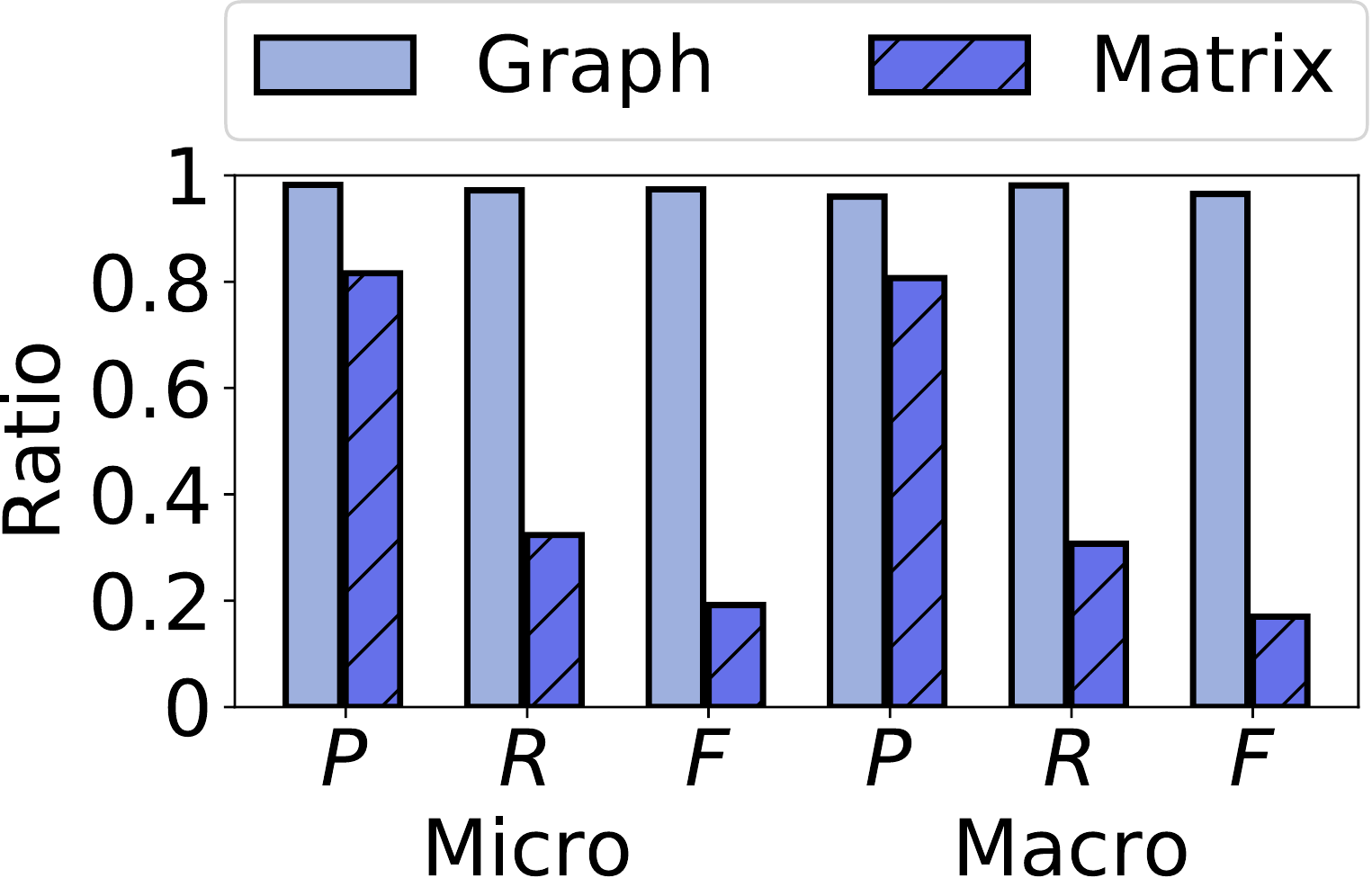}
        }
    	\caption{Graph modeling and \linep{} vs. matrix representation.}
    	\label{fig:stacked_embedding}
    	\vspace{-0.15in}
\end{figure}

We also evaluate the effectiveness of our bipartite graph modeling and \linep{} in \n{} compared to the case when \n{} is simply based on a matrix representation of RF signal samples, i.e., the matrix representation is directly used with the proximity-based hierarchical clustering. Here, for the matrix representation, the missing entries are filled with $-120$dBm, and each row is considered as an embedding of each sample. As shown in Figure~\ref{fig:stacked_embedding}, the matrix representation leads to quite poor classification performance, which clearly indicates the seriousness of the missing value problem. However, \n{} is inherently free of this problem and achieves outstanding performance, because \linep{} effectively captures the similarities and differences among the RF signal samples in the embedding space after their relationships are represented in the bipartite graph.

\subsection{System Parameter Evaluation}
\label{subsec:exp_micro}

\begin{figure}[t]
    \centering
    	\subfloat[Microsoft]{%
        \includegraphics[width=0.47\linewidth]{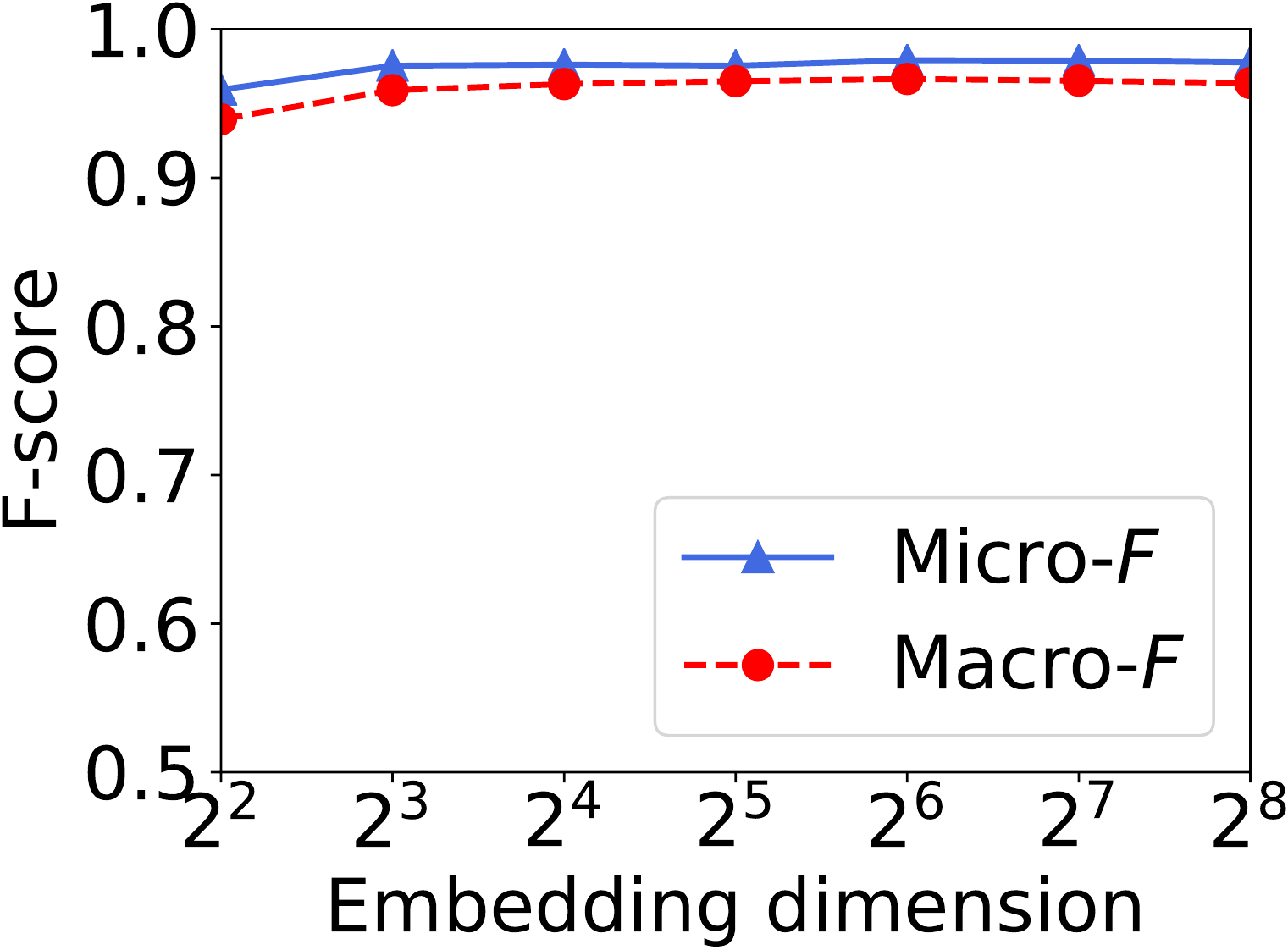}
        }
        \hspace{0.02in}
        \subfloat[Hong Kong]{%
        \includegraphics[width=0.47\linewidth]{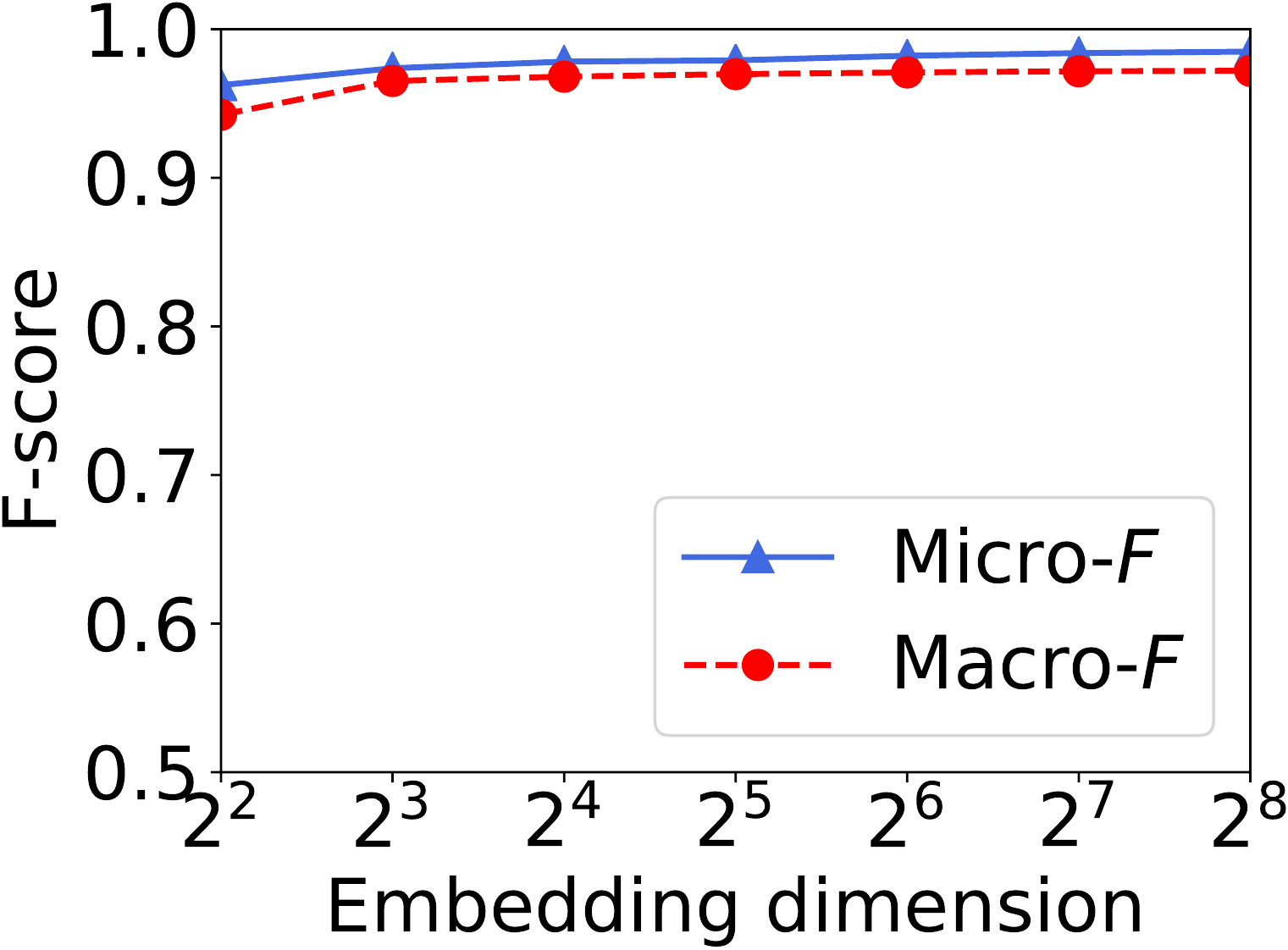}
        }
    	\caption{Insensitivity of \n{} to the choice of the embedding dimension.}
    	\label{fig:emb_dim}
    	\vspace{-0.1in}
\end{figure}

\n{} learns the embedding vector of each RF signal sample (each node of one type in the bipartite graph) for floor classification. It is thus important to check how the dimension of the embedding vector affects the system performance. Figure~\ref{fig:emb_dim} shows the $F$-scores when the dimension varies. \n{} consistently performs well regardless of the choice of the dimension. This indicates that \n{} does not require a careful choice of the dimension and its real deployment would be easier.

\begin{figure}[t]
    \centering
	\includegraphics[width=0.9\linewidth]{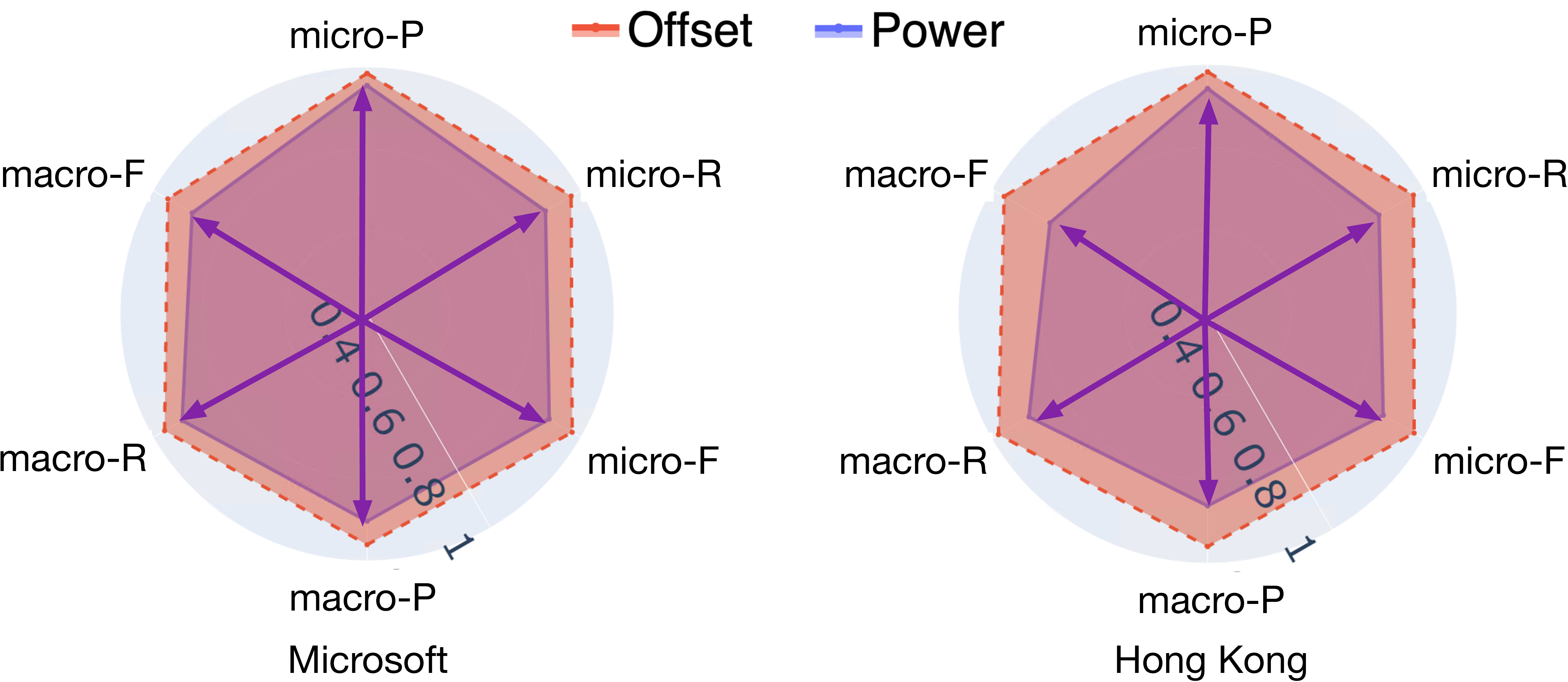}
    \caption{Impact of different weight functions on the performance.}
    \vspace{-0.2in}
	\label{fig:offset_power}
\end{figure}

When we construct a (weighted) bipartite graph, we need to ensure that the edge weights are non-negative for graph embedding. To achieve this, we define a proper weight function that adds a valid offset to all collected RSS values, i.e., $f(\text{RSS})\!=\!\text{RSS}\!+\!120$. Another choice of the weight function would be to use a function that converts each RSS value in dBm into the one in power, which is $g(\text{RSS})\!=\!10^{\text{RSS}/10}$. In Figure~\ref{fig:offset_power}, we present the comparison results between \n{} (with $f(\cdot)$) and \n{} that is used with $g(\cdot)$ instead of $f(\cdot)$. We see that the performance of using $f(\cdot)$ is substantially better than the one with $g(\cdot)$. The change in RSS values does not lead to much difference in $g(\cdot)$ compared to $f(\cdot)$, and thus the edge weights would be similar over different edges. Having similar edge weights makes the learned embeddings less effective. In other words, we observe that preserving the differences in RSS values is important to obtain high-quality node embeddings. We also tested different offset values and observed that the performance is more or less the same. We omit the results here for brevity.

\begin{figure}[t]
    \centering
    	\subfloat[Microsoft]{%
        \includegraphics[width=0.42\linewidth]{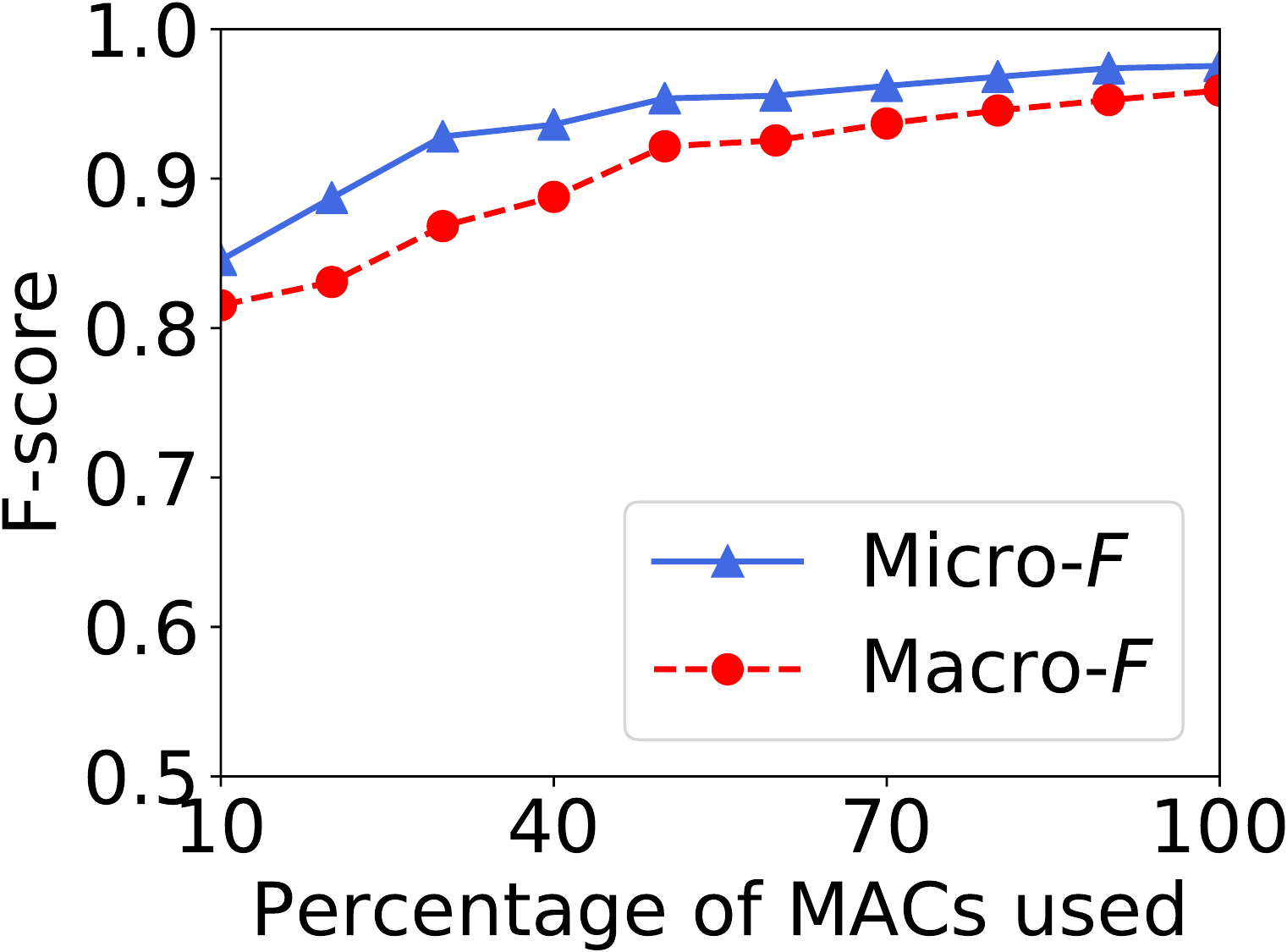}
        }
        \hspace{0.02in}
        \subfloat[Hong Kong]{%
        \includegraphics[width=0.42\linewidth]{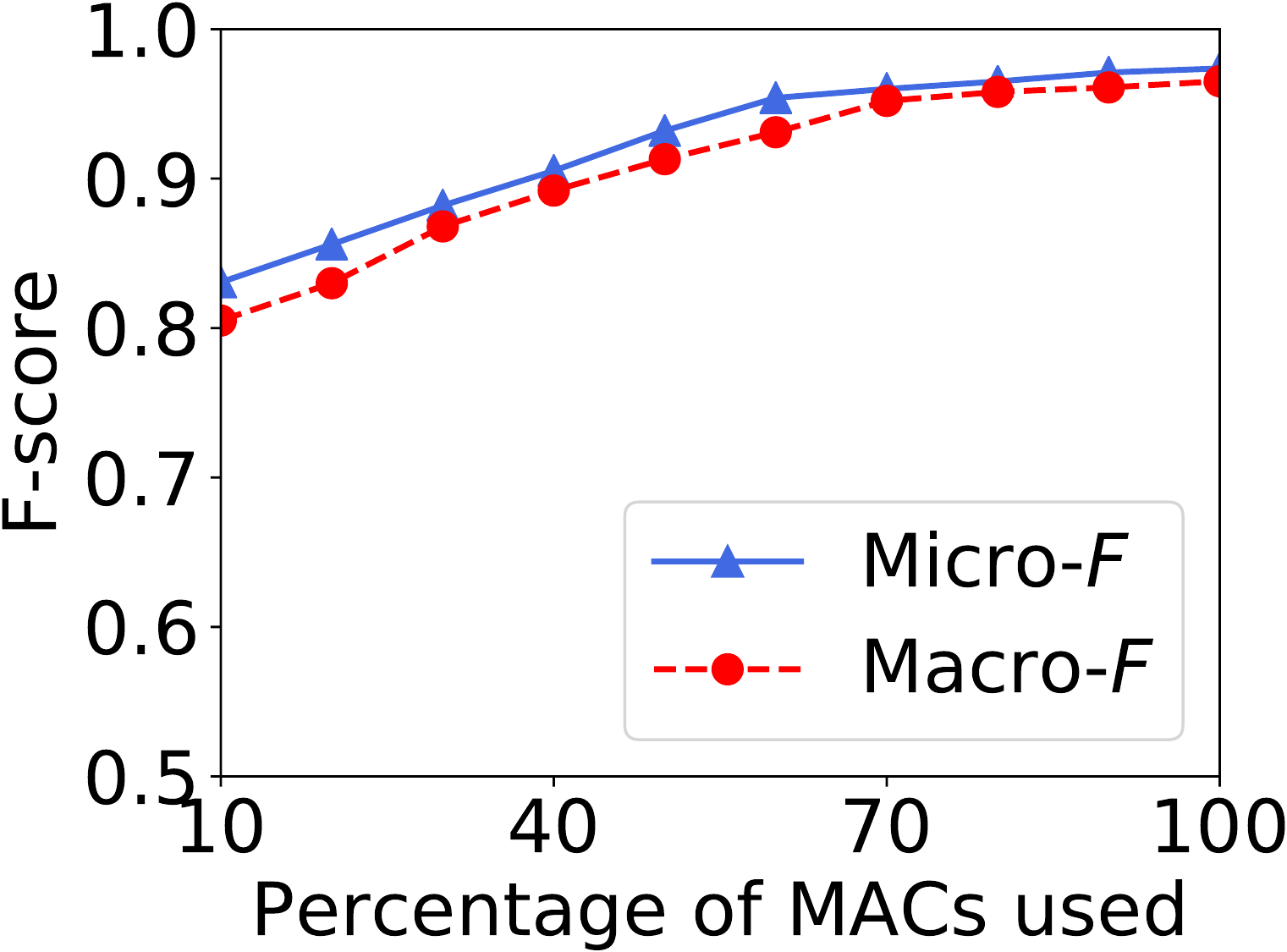}
        }
    	\caption{$F$-scores of \n{} with varying number of MACs available.}
    	\label{fig:ap_removal}
    	\vspace{-0.2in}
\end{figure}

To further validate the performance of \n{} when the ambient RF signals are from fewer APs (or fewer sensed MAC addresses), we consider the situation where a small fraction of MAC addresses are assumed to exist in the buildings. Figure~\ref{fig:ap_removal} shows the results of \n{} with varying size of the fraction. We see that even with 10\% of MAC addresses remaining on-site, \n{} can still achieve $F$ scores higher than 0.8, which demonstrates its robustness in sparse RF environments. With 30\% to 40\% of MACs available on-site, \n{} reaches $F$ scores higher than 0.9, which shows that \n{} can be readily deployed in practice.

\section{Conclusion}
\label{sec:conclude}
We have presented \n{} -- a novel graph embedding-based floor identification system, which consists of the novel bipartite graph modeling, high-quality graph embedding via \linep{}, and effective proximity-based hierarchical clustering. We have validated its performance on two large-scale datasets and demonstrated its superior prediction performance with only a few labeled samples over several state-of-the-art algorithms (by about 45\% in micro-$F$ score and 53\% in macro-$F$ score). We have also shown its various practical aspects for real deployment. 

\bibliography{ref}
\bibliographystyle{IEEEtran}

\end{document}